\documentclass[prd,showpacs,twocolumn]{revtex4}
\usepackage{url}
\usepackage{aas_macros,psfig,amsmath}
\usepackage{graphicx}
\begin{document}
\voffset 2cm

\def\bi#1{\hbox{\boldmath{$#1$}}}
\def\sun{\hbox{$\odot$}}
\def\farcs{\hbox{$.\!\!^{\prime\prime}$}}

\newcommand{\be}{\begin{equation}}
\newcommand{\ee}{\end{equation}}
\newcommand{\bea}{\begin{eqnarray}}
\newcommand{\eea}{\end{eqnarray}}

\newcommand{\band}[2]{{^{#1}\!{#2}}}
\newcommand{\lexp}{\mathop{\langle}}
\newcommand{\rexp}{\mathop{\rangle}}
\newcommand{\rexpc}{\mathop{\rangle_c}}

\newcommand{\lya}{Ly$\alpha$}
\newcommand{\lyaf}{Ly$\alpha$ forest}

\def\bi#1{\hbox{\boldmath{$#1$}}}

\def\affilmrk#1{$^{#1}$}
\def\affilmk#1#2{$^{#1}$#2;}

\def\ptonp{1}
\def\seattle{2}
\def\pton{3}
\def\apo{4}
\def\mit{5}
\def\tokyo{6}
\def\fnal{7}
\def\sussex{8}
\def\flagstaff{9}
\def\portsmouth{10}
\def\cambridge{11}
\def\psu{12}
\def\penn{13}
\def\osu{14}
\def\chicago{15}

\title{Cosmological parameter analysis including SDSS \lyaf\ and galaxy bias:
constraints on the 
primordial spectrum of fluctuations, neutrino mass, and dark energy}

\author{
Uro\v s Seljak \affilmrk{\ptonp},
Alexey Makarov \affilmrk{\ptonp},
Patrick McDonald \affilmrk{\ptonp},
Scott F. Anderson \affilmrk{\seattle},
Neta A. Bahcall \affilmrk{\pton},
J. Brinkmann \affilmrk{\apo},
Scott Burles \affilmrk{\mit},
Renyue Cen \affilmrk{\pton},
Mamoru Doi \affilmrk{\tokyo},
James E. Gunn \affilmrk{\pton}, 
\v{Z}eljko Ivezi\'{c} \affilmrk{\pton,\seattle},
Stephen Kent \affilmrk{\fnal},
Jon Loveday \affilmrk{\sussex},
Robert H. Lupton \affilmrk{\pton},
Jeffrey A. Munn \affilmrk{\flagstaff},
Robert C. Nichol \affilmrk{\portsmouth},
Jeremiah P. Ostriker \affilmrk{\pton,\cambridge},
David J. Schlegel \affilmrk{\pton},
Donald P. Schneider \affilmrk{\psu}, 
Max Tegmark\affilmrk{\penn,\mit},
Daniel E. Vanden Berk \affilmrk{\psu},
David H. Weinberg \affilmrk{\osu},
Donald G. York \affilmrk{\chicago}
}

\address{
\parshape 1 -3cm 24cm
\affilmk{\ptonp}{Physics Department, Princeton University, Princeton, NJ 08544,
USA} 
\affilmrk{\seattle}{Astronomy Department, University of Washington, Seattle, WA  98195, USA}
\affilmrk{\pton}{Princeton University Observatory, Princeton, NJ 08544,USA}
\affilmrk{\apo}{Apache Point Observatory, 2001 Apache Point Rd,Sunspot, NM 88349-0059, USA}
\affilmk{\mit}{Dept. of Physics, Massachusetts Institute of Technology,Cambridge, MA 02139, USA}
\affilmk{\tokyo}{Institute of Astronomy, School of Science, University of Tokyo, Japan}
\affilmk{\fnal}{Fermi National Accelerator Laboratory, P.O. Box 500, Batavia,IL 60510, USA}
\affilmk{\sussex}{University of Sussex, Sussex, UK}
\affilmk{\flagstaff}{U.S. Naval Observatory,Flagstaff Station, Flagstaff, AZ 86002-1149, USA}
\affilmrk{\portsmouth}{Institute of Cosmology and Gravitation, University of Portsmouth, Portsmouth, UK}
\affilmrk{\cambridge}{Institute of Astronomy, Cambrdige University, Cambridge, UK}
\affilmk{\psu}{Dept. of Astronomy and Astrophysics, Pennsylvania State University,University Park, PA 16802, USA}
\affilmk{\penn}{Department of Physics, University of Pennsylvania,
Philadelphia, PA 19104, USA}
\affilmk{\osu}{Department of Astronomy, Ohio State University,
Columbus, OH 43210, USA}
\affilmk{\chicago}{Department of Astronomy \& Astrophysics, University of
Chicago, Chicago, IL 60637, USA}
}

\date{\today. To be submitted to Phys. Rev. D.}

\begin{abstract}
We combine the constraints from the recent 
Ly-$\alpha$ forest analysis of the Sloan 
Digital Sky Survey (SDSS) and the
SDSS galaxy bias analysis
with previous constraints from SDSS galaxy clustering, the latest supernovae, 
and 1st year
WMAP cosmic microwave background anisotropies. 
We find significant improvements on all of the cosmological parameters 
compared to previous constraints, which
highlights the importance of combining \lyaf\ constraints with
other probes.
Combining WMAP and the \lyaf\
we find for the primordial slope $n_s=0.98\pm 0.02$.
We see
no evidence of running, $dn/d\ln k=-0.003\pm 0.010$, 
a factor of 3 improvement over 
previous constraints.  
We also find no evidence of tensors, $r<0.36$ (95\% c.l.).
Inflationary models predict the absence of running and 
many among them satisfy these constraints, particularly 
negative curvature models such as those
based on spontaneous symmetry breaking. A 
positive correlation between tensors and primordial slope disfavors chaotic 
inflation type models with steep slopes: 
while the $V \propto \phi^2$ model 
is within the 2-sigma contour, 
$V \propto \phi^4$ is outside the 3-sigma contour. 
For the amplitude we find $\sigma_8=0.90\pm 0.03$ from the \lyaf\ and WMAP 
alone. 
We find no evidence 
of neutrino mass: for the case of 3 massive
neutrino families with an inflationary prior, 
$\sum m_{\nu}<0.42$eV and the mass of lightest neutrino 
is $m_1<0.13$eV at 95\% c.l. 
For the 3 massless + 1 massive neutrino case we 
find $m_{\nu}<0.79$eV for the massive neutrino, excluding at 95\% c.l. 
all neutrino mass solutions compatible with 
the LSND results. 
We explore dark energy constraints 
in models with a fairly general time dependence of dark energy 
equation of state, finding $\Omega_{\lambda}=0.72\pm 0.02$,
${\rm w}(z=0.3)=-0.98^{+0.10}_{-0.12}$, the latter changing 
to ${\rm w}(z=0.3)=-0.92^{+0.09}_{-0.10}$  if tensors are 
allowed. 
We find no evidence for variation of the equation of state with redshift,
${\rm w}(z=1)=-1.03^{+0.21}_{-0.28}$.
These results rely on the current understanding of the \lyaf\ and other 
probes, which 
need to be explored further both observationally and theoretically, 
but extensive tests reveal no evidence of inconsistency among different
data sets used here.

\end{abstract}



\maketitle

\setcounter{footnote}{0}

\section{Introduction}

Many different cosmological observations over the past decade have helped build 
what is now called the standard cosmological model. These observations
suggest that the universe is spatially flat, contains 
baryons, dark matter and dark energy. The primordial spectrum of fluctuations 
is approximately scale invariant and initial fluctuations are Gaussian 
and adiabatic. 
This standard cosmological model can be described in terms of only 
a few parameters, which explain a large number of observations, such as
the cosmic microwave background (CMB), galaxy clustering, 
supernova data, Hubble parameter determinations,
and weak lensing. The latest results 
come from Wilkinson Microwave Anisotropy Probe (WMAP) CMB measurements 
\cite{2003ApJS..148....1B,2003ApJS..148..135H,2003ApJS..148..161K}, 
Sloan Digital 
Sky Survey (SDSS) and Two degree Field (2dF) galaxy clustering analyses 
\cite{2004ApJ...606..702T,2004ApJ...607..655P,2001MNRAS.327.1297P}, and 
from the latest Supernovae type Ia (SNIa) data \cite{2004ApJ...607..665R,2003ApJ...598..102K}. 

While the standard model is observationally well justified, many theoretical 
models predict that there should be observable deviations from it. 
Perhaps the best motivated among these are the predictions of how the
universe was seeded by initial fluctuations. 
The standard paradigm is inflation, which predicts that the fluctuations 
should 
be almost, but not exactly, scale invariant \cite{2000cils.conf.....L}.  
A typical deviation for the slope of the primordial perturbations 
is predicted to be of order of a few parts in a hundred away from its 
scale invariant value $n_s=1$ and could be of either sign. 
This should be observable with 
high precision cosmological observations. Despite tremendous progress over 
the past couple of years the current constraints do not yet distinguish 
between different inflationary models 
\cite{2003ApJS..148..213P,2004PhRvD..69j3501T}. 
Alternative models also predict 
deviations from scale invariance similar to 
inflation \cite{2003PhRvL..91p1301K}. 
Another prediction of these models is that the rate of change of slope 
with scale is rather small, $\alpha_s=dn_s/d\ln k \sim (n_s-1)^2 \sim 10^{-3}$, 
which should not be observable in the near future. 
A third prediction that can distinguish among the different models 
is the amount of tensor perturbations they predict. Some models 
predict no detectable 
tensor contribution \cite{2000cils.conf.....L,2001PhRvD..64l3522K}, 
while other models predict a tensor contribution to the 
large scale CMB anisotropies comparable to that from scalars. 
It is clear that determining the shape and amplitude of the scalar and tensor
primordial power spectra will be one of the key tests of various models 
of structure formation. 

Current observational constraints on the primordial power spectrum 
are mostly limited to scales larger 
than 10$h^{-1}$Mpc. There are various reasons for this: 
CMB fluctuations are damped on small scales and their detection 
would require high resolution,
low noise detectors, which are only now being built. Even with 
sufficient signal-to-noise and angular resolution there may be secondary 
anisotropies that may contaminate the signal from primary anisotropies. 
On small scales, matter undergoes strongly nonlinear evolution, 
which erases the 
initial spectrum of fluctuations and prevents 
galaxy clustering and 
weak lensing surveys from extracting this information.
On the other end, 
the largest observable scale is 
the horizon scale seen by CMB fluctuations. 
The small number of available modes on the sky prevents one from accurately 
determining the primordial spectrum on these scales from the CMB. 
The largest scales probed by galaxy clustering are even smaller. 
As a result, the primordial power spectrum is currently probed 
over a relatively narrow range of scales and the shape of the
primordial power spectrum cannot be accurately determined. 

To improve these constraints one should determine the 
fluctuation amplitude  on smaller scales. Nonlinear evolution prevents 
one from obtaining useful information at $z=0$, so one must 
look for probes at higher redshift. 
Of the current cosmological
probes, the Ly-$\alpha$ forest -- the absorption observed in quasar spectra
by neutral hydrogen in the intergalactic medium (hereafter IGM) --
has the potential to give the most precise information on small scales 
\cite{1998ApJ...495...44C}.
It probes fluctuations down to megaparsec scales at redshifts between 
2-4, so nonlinear evolution, while not negligible, has not erased 
all of the primordial information. 

In this paper we combine CMB/LSS constraints with the new 
analysis of the Ly-$\alpha$ forest from SDSS data 
\cite{2004astro.ph..5013M}. 
The Sloan Digital Sky Survey \citep{2000AJ....120.1579Y} uses
a drift-scanning
imaging camera \citep{1998AJ....116.3040G} and a 640 fiber, double
spectrograph on a dedicated 2.5
m telescope. 
The SDSS data sample in data release two \citep{2004astro.ph..3325A}
consists of more than 3000 QSO spectra with $z>2.2$, 
nearly two orders of magnitude
larger than previously available 
\cite{2002ApJ...581...20C,2000ApJ...543....1M,2003astro.ph..8103K}. 
This large data set 
allows one to determine the amplitude of the flux power spectrum to better than 
1\%. Theoretical analysis of this flux power spectrum shows that at the 
pivot point k=0.009 s/km in velocity coordinates, 
which is close to k=1h/Mpc in comoving coordinates for 
standard cosmological parameters, the power spectrum amplitude 
is determined to about 15\% and the slope to about 0.05,
with the error budget dominated by uncertainties in theoretical modelling
\cite{McDonald04b,McDonald04c}. 
This is an accuracy 
comparable to that achieved by WMAP. 
More importantly, it is 
at a much smaller scale, so combining the two leads to a significant 
improvement in the constraints on primordial power spectrum shape over what 
can be achieved from each data set individually. 

A second theoretical prediction where the basic cosmological model is 
expected to 
require modifications is that neutrinos have mass. Atmospheric mixing 
and solar neutrino results suggest that the  
total minimum neutrino mass is about 
0.06eV \cite{1998PhRvL..81.1158F,2004hep.ex..4034A,2001PhRvL..87g1301A}. 
These 
observations are only sensitive to relative neutrino mass 
differences and not to the absolute neutrino mass itself. 
Cosmology on the other hand can weigh neutrinos directly. 
Massive neutrinos slow down the growth of structure on small scales and
modify the amplitude and shape of the matter power spectrum. They also 
modify the CMB power spectrum. If one measures both the CMB and matter power 
spectra with high precision across a wide range of redshifts and 
scales then one can determine the neutrino mass with high accuracy \cite{1998PhRvL..80.5255H}. 
The question of neutrino mass is also interesting in light of recent 
Los Alamos Liquid Scintillator Neutrino Detector (LSND)
experimental results, which, if taken at a face value, 
suggest $m_{\nu}>0.9$eV 
\cite{1996PhRvL..77.3082A,2003JCAP...05..004H,2004PhLB..581..218P}, 
which should be observable by 
cosmological neutrino weighing. 

A third theoretical prediction of departures from the standard model, and 
one whose consequences would be particularly far reaching, is that 
dark energy is not simply a cosmological constant introduced 
already by Einstein, but something more complicated and dynamical in 
nature. In the case where dark energy is a scalar field one 
would expect that it has a kinetic energy term in addition to the
potential term, which modifies its equation of state. This is expected to 
evolve with time, but theoretical predictions are rather uncertain and 
are suggestive at best. A change in equation of state 
changes both the rate of growth of structure and the angular size of 
the acoustic horizon in the CMB. As a result these changes can be observed both 
through the CMB and by comparing the growth of structure at different 
redshifts. 

Many different methods have been discussed in the literature on how to 
improve the current constraints from methods such as supernovae type Ia (SNIa),
CMB, weak lensing, and cluster abundances.
One method to constrain the nature of dark energy 
that has not attracted much attention, yet has the potential 
to produce results on a relatively short time scale, is comparing
measurements of amplitude of fluctuations at high redshift from the
\lyaf\ and CMB to that at low redshift from galaxy 
clustering. 
Dark energy affects the rate of growth of structure, especially 
for $z<1$ where dark energy is dynamically important. 
In this paper we combine WMAP and SDSS \lyaf\
measurements at high 
redshifts, where 
dark energy is expected to be negligible, with the amplitude determination 
at $z=0.1$ from the SDSS galaxy bias analysis \cite{2004astro.ph..6594S}. 
In general, galaxy clustering is believed to be proportional to 
matter clustering on large scales 
up to a constant of proportionality. This constant, the so called 
bias, is a free parameter that cannot be determined from the clustering 
analysis itself. There are many different methods for how to determine the
bias and thus the amplitude of matter fluctuations such as 
redshift space distortions \cite{1999MNRAS.310.1137H,2004ApJ...606..702T}, 
the bispectrum \cite{2002MNRAS.335..432V}, or weak lensing 
\cite{2002ApJ...577..604H,2003astro.ph.12036S}, but the current 
constraints are weak. A recent analysis of the luminosity dependence of 
galaxy clustering \cite{2004ApJ...606..702T}, combined with a determination 
of the halo mass 
distribution for these galaxies, provides a new constraint on the 
bias and amplitude of fluctuations in SDSS data \cite{2004astro.ph..6594S}. 

One difference of the current paper 
in comparison with previous analyses of this type is that 
we present 68.32\%, 95.5\% and 99.86\% confidence intervals 
(we denote these the 1, 2, and 3-$\sigma$ intervals, 
but note that they do not depend on the
assumption of Gaussianity in the error distribution) on all the parameters 
(or 95\% and 99.9\% confidence level upper 
limits in the case of no detections). 
Sometimes the 3-$\sigma$ intervals can be 
significantly different from 3 times the corresponding 1-$\sigma$ intervals. 
This can happen if there are degeneracies in the data that 
appear to be broken at 1-$\sigma$, but that the 2 or 3 $\sigma$ 
contours allow. In this case the 3-$\sigma$
constraints are weaker than the corresponding 
1-$\sigma$ intervals would suggest. The opposite can happen as well, 
especially if there is a natural boundary that the parameter cannot 
cross (such as a parameter being positive definite). 
More generally, presenting 1-$\sigma$ 
contours alone is not very 
meaningful, since whatever is within 1-$\sigma$ is essentially a good fit 
to the data. One can argue that the goal of observations
is to exclude regions of parameter 
space and this is much better represented by reporting 2 and 3-$\sigma$ 
contours than the best fit value and its 1-$\sigma$ range.  

Another issue that we address in detail is the robustness of the 
constraints against the number of parameters one is exploring.  
Sometimes the constraints 
change significantly if new parameters are added to the mix because 
these new parameters are degenerate with parameters one is 
interested in.
However, often the quality of the fit is not improved at all and 
moreover
these new parameters may not be well motivated from 
the perspective of fundamental theories or other considerations. 
In this case one is entitled to 
adopt an Occam's razor argument against the introduction of these
parameters in the 
estimation. 
To some extent this is always a subjective 
procedure, since what is natural for one person may not be for 
someone else. It has also been argued that one should pay a 
penalty for each new parameter that is introduced which does not 
improve the quality of the fit \cite{2004astro.ph..1198L}. 
However, this procedure is also 
poorly defined and there is no unique choice for the penalty. 
In this paper we explore both the solutions with 
the minimum number of parameters as well as with several 
additional parameters. We believe that there is merit to the approach 
which parametrizes the constraints with as few parameters as possible, so 
our main results are given for this case. However, one also wants to know 
how robust and model independent are the constraints, which we 
explore by adding several additional parameters to the analysis. 

The outline of this paper is as follows. We first present the method, 
then our basic results in several tables and then discuss them in detail. 
We focus particularly on the question of how have the new results improved
upon the previous constraints and how robust are the conclusions upon 
removing one or more of the data ingredients. The latter is particularly 
interesting in light of possible systematic effects that may be 
present both in the new analyses of \lyaf\ and bias as well as in 
previous analyses of WMAP, SDSS galaxy clustering, and SNIa. 

\section{Method} 

We combine the constraints from the SDSS Ly-$\alpha$ 
forest \cite{2004astro.ph..5013M}
with the SDSS galaxy clustering analysis \cite{2004ApJ...606..702T}, 
SDSS bias analysis 
\cite{2004astro.ph..6594S}, and CMB power spectrum 
observations from WMAP 
\citep{2003ApJS..148....1B,2003ApJS..148..135H,2003ApJS..148..161K}. We
verified that including CBI, VSA, and ACBAR
\citep{2004astro.ph..2359R,2004astro.ph..2466R,2002astro.ph.12289K} makes 
very little
difference in the final results and we do not include them in the current analysis.
Similarly, we verified that including the latest 2dF power spectrum analysis 
\cite{2001MNRAS.327.1297P} in addition to SDSS does not make much difference, so we do not 
include those constraints either. 
We could have used 2dF constraints instead of SDSS, 
but we chose not to because 
for 2dF the bias constraints are somewhat 
weaker \cite{2002MNRAS.335..432V} and we 
would like to have an independent verification of results that 
use the 2dF bias \cite{2003ApJS..148..175S}.  
We will thus refer to CMB constraints
as WMAP, to LSS/galaxy clustering constraints as SDSS-gal, to 
SDSS bias constraints as SDSS-bias and to SDSS Ly-$\alpha$ forest 
constraints as SDSS-lya. We have added earlier \lyaf\ 
constraints in a weak form 
\cite{2000ApJ...543....1M,2001ApJ...562...52M}, 
which have a small, but not negligible effect. 
We do not include more recent \lyaf\ constraints 
\cite{2002ApJ...581...20C,2003astro.ph..8103K} since there are 
signs of systematic discrepancy and/or underestimation of errors
when compared to SDSS \lyaf\ data \cite{2004astro.ph..5013M}. 
To this we add the latest supernova 
constraints as given in \cite{2004ApJ...607..665R}. 
We do not use this full combination in 
all calculations, since we want to emphasize what the new 
constraints bring to the mix and we want to explore the 
sensitivity of the constraints to individual data sets. For example, for 
the investigation of the shape of 
the primordial power spectrum we perform the analysis using WMAP+SDSS-lya
alone and show that this combination in itself suffices to constrain the 
running by a factor of 3 better than combining everything else together. 
We also perform several analyses by dropping one of the constraints
and explore the robustness of the conclusions. For example, we explore the 
constraints on the dark energy equation of state with and without SNIa 
and with and without SDSS-bias and SDSS-lya. 

Our implementation of
the Monte Carlo Markov Chain (MCMC) method
\citep{2003MNRAS.342L..79S} uses
CMBFAST 
\cite{1996ApJ...469..437S}
version 4.5.1\footnote{available at cmbfast.org}, 
outputting
 both CMB spectra and the corresponding matter power spectra $P(k)$.
We evolve all the matter power spectra to a high $k$ using CMBFAST and
we do not employ any analytical approximations. We output the transfer 
functions at the redshifts of interest, between 2-4 for SDSS-\lyaf\ and 0.1
for SDSS-gal. 
Note that for massive neutrinos the high precision (HP) 
option must be used to achieve sufficient accuracy in the transfer function. 

A typical run is based on 16-24 independent chains,
contains 50,000-200,000 chain elements and requires several days
of running on a computer cluster in a serial mode of CMBFAST.
The acceptance rate was of order 30-50\%,
correlation length 10-30
and the effective chain length of order 3,000-20,000 (see 
\cite{2004PhRvD..69j3501T} for definitions of these terms).
In terms of Gelman and Rubin $\hat{R}$-statistics \citep{gelman92} we find the
chains are sufficiently converged and mixed, with $\hat{R}<1.05$,
significantly more conservative than the recommended value $\hat{R}<1.2$.

Our most general cosmological parameter space is
\be
\bi{p}=(\tau,\omega_b,\omega_m,\sum m_{\nu},\Omega_{\lambda}, {\rm w}, \Delta^2_{\cal R}, n_s,
\alpha_s,r),
\ee
where $\tau$ is the optical depth, $\omega_b=\Omega_bh^2$, where $\Omega_b$ 
is baryon density in 
units of the critical density and $h$ is the 
Hubble constant in units of 100km/s/Mpc, 
$\omega_m=\Omega_mh^2$ where $\Omega_m$ is matter density in units of 
the critical density, 
$\sum m_{\nu}$
is the sum of 
massive neutrino masses (assuming either 3 degenerate neutrino families 
or 1 massive neutrino family in addition to 3 massless),
$\Omega_{\lambda}$ is the dark energy density today and $w$ its equation
of state (which is in general time dependent). 
Our pivot point for the primordial power spectrum parameterization 
is at $k_{{\rm pivot}}=0.05/$Mpc and we expand 
the primordial power spectrum at that point, defining the amplitude 
of curvature perturbations $\Delta^2_{\cal R}$, slope 
$n_s$, and its running $\alpha_s=dn_s/d\ln k$. The choice of the pivot 
point is somewhat arbitrary, but is meant to represent the scale somewhere 
in the middle of the observational range. In this case the largest scales are
probed by the CMB ($k \sim 10^{-3}$/Mpc) and the smallest scales are 
probed by the \lyaf\
($k \sim 1$/Mpc). In addition, this scale has been (arbitrarily) chosen 
as a pivot point 
in CMBFAST and has been used by previous analyses, which facilitates the 
comparison. Note that there is no Hubble parameter $h$ in the definition of 
the pivot point: if
CMB data are used there is no advantage in defining the 
scale by taking out the Hubble constant, 
unlike the case of galaxy clustering and \lyaf.  

We parametrize tensors in terms of their amplitude $\Delta_h^2$, and
define the ratio relative to 
scalars as  $r=T/S=\Delta_h^2/\Delta^2_{\cal R}$. 
This is also defined at the pivot point $k=0.05$/Mpc,
just as for the scalar amplitude, slope
and running.
We fix the tensor
slope $n_T$ using $r=-8n_T$.
We do not allow for non-flat
models, since curvature is
already tightly constrained by CMB and
other observations
\cite{2003ApJS..148..175S}.
In addition, we will be testing particular classes of models, such as 
inflation, which predict $K=0$. 
For the more general models, such as those with freedom in the dark energy
equation of state, relaxing this assumption can lead to a significant
expansion of errors \cite{2004PhRvD..69j3501T}. 
We are therefore testing a particular class of
inflation inspired models with $K=0$
and not presenting model independent constraints on the equation of state.
Note that this assumption is implicit in most of the constraints 
published to date, including those from the SNIa teams, which often assume 
a CMB prior on $\Omega_m$ \cite{2004ApJ...607..665R}. This prior is affected by the choice 
of parameter space one is working in and a self-consistent treatment is 
required. CMB constraints on $\Omega_m$ using an analysis where 
the equation of state or curvature are not varied need not equal 
those where these are varied. 
We follow the WMAP team in imposing a $\tau<0.3$ constraint. Upcoming
polarization data from WMAP will allow a verification of this prior.

From this basic set of parameters we can obtain constraints on
several other parameters, such as the baryon and matter densities
$\Omega_b$ and $\Omega_m$, Hubble parameter $h=H_0$/(100km/s/Mpc) 
and amplitude of fluctuations $\sigma_8$.
Since we do not allow for curvature we have
$\Omega_{\lambda}=1-\Omega_m$ and we use $\Omega_m$ in
all tables. In fact, our primary parameter
is the angular scale of the acoustic horizon, which is tightly
constrained by the CMB.
Similarly, although we use $\Delta_{\cal R}^2$ as the primary parameter in
the MCMC we
present the amplitude in terms of the more familiar $\sigma_8$.
In addition to the cosmological parameters above we also keep track of 
several parameters related to the specific tracers, described below. 

\subsection{CMB analysis}
For the CMB we use the 1st year likelihood routine provided by WMAP
\cite{2003ApJS..148..135H,2003ApJS..148..195V}, but replace $l<12$ analysis with the
corresponding full
likelihood analysis as given in \cite{2004astro.ph..3073S}.
This is important for 
the running of the spectral index constraints. As shown in 
\cite{2004astro.ph..3073S},
exact analysis increases 
errors on low multipoles compared to the original WMAP analysis, 
which leads 
to less stringent constraints on running: 
it is typically increased
by one standard deviation
away from its negative value toward zero, i.e. toward 
the no running solution. We 
find a similar effect in our analysis 
when combined with Ly-$\alpha$ forest analysis.  

\subsection{Galaxy clustering}
We use the SDSS galaxy clustering constraints on the galaxy power spectrum 
for $k<0.2h$/Mpc \cite{2004ApJ...606..702T}. 
We use a
linear to nonlinear mapping of the matter power spectrum using expressions
given in \cite{2003MNRAS.341.1311S}.
The main nuisance 
parameter is the linear bias of $L_*$ galaxies, $b_*$, 
which relates the galaxy power spectrum 
to that of dark matter, $P_{\delta_{\rm g}}(k)=b_*^2P_{\delta_{\rm dm}}(k)$,
where $\delta_{\rm g}$ and $\delta_{\rm dm}$
are the galaxy and dark matter density fluctuations, respectively,
and $P(k)$ is their power spectrum. 

The luminosity dependence of galaxy 
bias provides additional cosmological constraints \cite{2004astro.ph..6594S}. 
Observations show that bias is relatively constant for 
galaxies fainter than $L_*$ and is rapidly increasing for 
brighter galaxies \cite{2004ApJ...606..702T}. 
Theoretical and simulation 
predictions of halo bias 
\cite{1998ApJ...503L...9J,1999MNRAS.308..119S,2004astro.ph..3698S} show a 
similar 
dependence of bias on halo mass, with the transition occurring at the 
so called nonlinear mass, corresponding to the mass within a sphere where 
the rms fluctuation level is 1.68. The value of the nonlinear mass 
depends on cosmological parameters such as the amplitude and shape of 
the power spectrum, as well as the matter 
density. 
A measurement of the halo mass 
distribution for a given luminosity class is possible using a
weak lensing analysis around these galaxies, which traces the dark matter 
distribution directly. This allows a theoretical determination 
of galaxy bias for a given cosmological model. Only those models 
for which the theoretical predictions agree with the observations 
in all luminosity bins
are acceptable. This places strong constraints on cosmological models. 
This constraint is not directly determining the amplitude of fluctuations
and bias, because both the theoretical predictions and observationally 
inferred values of bias change in a similar way. 
However, the data suggest that for $L_*$, where statistical errors 
are smallest,
the predicted bias value is lower than 
the observed one for standard cosmology $\Omega_m=0.3$ and $n_s=1$. 
Lowering $\Omega_m$ or $n_s$ reduces the nonlinear mass and increases
theoretically predicted bias, bringing it into a better agreement 
with observations. Additional constraints come from the dependence of
bias on luminosity, which is constraining the amplitude 
of fluctuations.  
The method is fairly robust in the sense that even appreciable changes in 
halo mass determination do not change the bias predictions significantly.
The analysis is performed using the bias likelihood 
code as given in \cite{2004astro.ph..6594S}. 

\subsection{\lyaf}
Reference \citep{McDonald04b} describes in detail our method for obtaining 
the \lyaf\ contribution to $\chi^2$ for any cosmological model.
Rather than attempting to invert $P_F(k)$ to obtain the
matter power spectrum, we compare the theoretical $P_F(k)$
directly to the observed one.
In observationally favored models, the Universe is effectively
Einstein-de Sitter at $z>2$,
so the cosmology information relevant to the Ly-$\alpha$ forest is completely
contained within $P_L(k)$ measured in velocity
units.  
For any given model in the MCMC chain 
we compute the matter power spectrum in velocity units and interpolate  
from a grid of cosmological simulations covering a broad 
range of values to obtain predictions of the flux power spectrum. 
We compare these to the measured SDSS flux power 
spectrum to derive the likelihood of the model given the data. 

The \lyaf\ contains several nuisance parameters which we are not interested
in for the cosmological analysis, although some of them are of interest for 
studies of IGM evolution. In the standard picture of the Ly-$\alpha$ forest
the gas in the IGM is in ionization equilibrium.  The
rate of ionization by the UV background balances the rate of
recombination of protons and electrons.
The recombination rate depends on the temperature of the gas, which
is a function of the gas density.
The temperature-density relation can be parameterized by an amplitude,
$T_0$, and a slope $\gamma-1=d\ln T/d\ln \rho$.
The uncertainties in the intensity of the UV background, the mean
baryon density, and other
parameters that set the normalization of the relation between optical
depth and density can be combined into one parameter: the mean transmitted
flux, $\bar{F}(z)$.
The parameters of the gas model, $T_0$, $\gamma-1$, and $\bar{F}$,
must be marginalized over when computing constraints on cosmology.
They are a function of redshift. Our model for the redshift evolution 
of $\bar{F}$, $T_0$, and $\gamma$ is explained in detail 
in \cite{McDonald04b}. 
We also add additional nuisance parameters such as the filtering 
length $k_F$ \cite{1998MNRAS.296...44G} and parameters that characterize 
various 
physical effects \cite{McDonald04c}, described in more detail below. 
This gives rise to a number of additional nuisance parameters.

Each time there are nuisance parameters that one is not interested in there
are two approaches that one can take. One can either keep these parameters 
as independent and add them to the MCMC chain or one can marginalize over them 
for each set of cosmological models. The advantage of the first approach is 
that at the end one can extract the best fit values of these parameters and 
their 
correlations with other cosmological or nuisance parameters, in case one is 
interested in these. The disadvantage is that increasing the number of 
dimensions of the MCMC decreases the acceptance rate of the chains, increasing 
the computational time. Another disadvantage is that in many dimensions the MCMC
approach often does not find the global minimum, which is of interest if one 
wants to assess the improvement in $\chi^2$ with the addition of new parameters. 

The second approach is marginalization over the nuisance parameters. 
We implement 
it by maximizing the likelihood (minimizing $\chi^2$) over the phase space 
of these parameters for each cosmological model. 
The computational efficiency of this approach depends on 
the problem at hand and numerical implementation. 
In our case we find that the computational time increase
is comparable to the penalty paid in the first approach due to the lowered 
efficiency of the MCMC sampler, so there is no numerical advantage in 
using one over the other. We decided to use the latter approach because 
we would like to be able to interpret the minimum $\chi^2$ values between 
different chains: 
we have 
found that the marginalization approach gives a minimum $\chi^2$ within 
unity of 
the global minimum for the chain lengths we adopt, while the approach of
working in 40-dimensional parameter space gave minimum $\chi^2$ values in 
our MCMC chains that 
were often significantly higher than the actual 
global minimum. This is expected since
the likelihood function is shallow around the maximum and the large phase space
volume of 20 additional dimensions wins over
the penalty induced by $\exp(-\Delta \chi^2)$ 
for small $\Delta \chi^2$. 
We have 
verified that both approaches lead to the same probability distributions of 
cosmological parameters, so this choice is not important for the MCMC
distributions themselves. 

More details of the \lyaf\ likelihood module are described in 
\cite{McDonald04b}. 
The simulations cover the plausibly allowed
range of $\bar{F}$, $T_0$, $\gamma-1$, $k_F$, $\Delta^2(k_{\rm eff})$, $n_{\rm eff}$,
and $dn_{\rm eff}/d\ln k$.
Simulations with several box and grid sizes are used to guarantee
convergence, which is verified by detailed convergence studies
on smaller box simulations.
The grid is based on hydro-particle mesh simulations \cite{1998MNRAS.296...44G},
but these are explicitly calibrated using fully hydrodynamic simulations
\citep{1994ApJ...437L...9C,2002astro.ph..3524C}.
The simulation results are combined in an interpolation code that
produces $P_F(k)$ for any relatively smooth (CDM-like) input
$P_L(k)$, $\bar{F}$, $T_0$, and $\gamma-1$. We also marginalize over the 
filtering scale $k_F$, which is related to the gas Jeans scale, where 
pressure balances gravity, but 
depends on the full gas temperature history since reionization  
rather than just the 
instantaneous temperature $T_0$ \cite{1998MNRAS.296...44G}. 

There are several possible systematic effects in the \lyaf\ 
that have been investigated 
in \cite{McDonald04c}. The most important effect, that from damped systems, 
can be 
reliably removed using the existing constraints on the abundance of 
damped systems. It leads to an increase in slope by 0.06. 
We find no evidence of other effects, such as fluctuations 
in the UV background or galactic winds. The former effect is constrained by 
the expected rapid evolution of the attenuation length with redshift, which 
would cause the effect to be more significant at high redshift. While 
current models 
of galactic winds produce no significant effect on the \lyaf\ flux power 
spectrum \citep{McDonald04c}, these need to be explored further. 
The fact that the effective curvature of the matter 
power spectrum derived solely from \lyaf\ analysis agrees with the 
expected value \cite{McDonald04b}
provides a constraint on any additional contamination. An independent 
constraint is provided by the consistency of the matter power spectrum 
results as a function of redshift over the range $2<z<4$ \cite{McDonald04b}.  
Neither of these arguments are conclusive and we find examples of 
systematic effects that can escape one or the other test.
Additional analyses, such as correlations of the \lyaf\ with galaxies 
\cite{2003ApJ...584...45A}
and quasars \cite{2003astro.ph..7563S,2003astro.ph.10890C}, as well as 
a bispectrum analysis \cite{2003MNRAS.344..776M}, will be able to test 
further the current models. 

\section{Results}
The basic results for many different MCMC runs are given in tables 1-4.
We give results for many different parameter combinations and different 
experiment combinations, with the purpose of assessing the robustness 
of constraints on both the data and parameter space. 
For most of the parameters we quote the median value (50\%),
[15.84\%,84.16\%] interval ($\pm 1\sigma$), [2.3\%,97.7\%] interval
($\pm 2\sigma$) and [0.13\%,99.87\%] interval
($\pm 3\sigma$). These are calculated from the cumulative one-point
distributions of MCMC values for each parameter and do not depend on the
Gaussian assumption. For the parameters without a detection we quote
a 95\% confidence upper limit and a 99.9\% confidence upper limit. 
We have found that our MCMC gives a reliable estimate of 3-sigma 
contours for one-dimensional projections. The corresponding 2-d 
projections are however very noisy and we do not plot 3-sigma 
contours in our 2-d plots. 

All of the restricted
parameter space fits are acceptable based on $\chi^2$ values, starting
from the basic 6-parameter model 
$\bi{p}=(\tau,\omega_b,\omega_m,\Omega _{\lambda}=1-\Omega_m, \Delta^2_{\cal R},n_s)$.
We denote this as 6-p in the tables. 
Introducing additional parameters such as 
tensors, running, equation of state, or neutrino mass
does not improve the fits.
We do not report the values of nuisance parameters such as the galaxy bias or 
\lyaf\ mean flux, temperature-density relation, or filtering length. Some
of these are discussed elsewhere \cite{2004astro.ph..6594S,McDonald04b}. 
When comparing the improvements over previous analyses we try to compare
the results to 
our own MCMC analysis of previous data. This is because small changes in 
the treatment, such as assumed priors,
can affect the parameters and so the constraints between 
different groups are 
not directly comparable. When comparing our analysis 
to \cite{2004PhRvD..69j3501T} 
we find in general a very good agreement between the two, even though 
our MCMC implementation 
is independent. Our primary goal is to determine how 
much the new data improve over the previous situation and to 
answer this it is best to perform identical analyses with and without 
the new data. 
Below we discuss the results from these tables in more detail.

\begin{table*}
\noindent
{\footnotesize
Table 1: Constraints on basic 6 parameters and tensors. Median value, 
$1\sigma$, $2\sigma$ and $3\sigma$ 
intervals on cosmological parameters combining WMAP, SDSS galaxies
(gal), SDSS bias (bias), SDSS \lyaf\ (lya)
and SNIa (SN) data as derived from the MCMC analysis. In each case 
we list individual data sets. Note that WMAP is included in all 
the chains. 
In the absence of a detection we give 95\% upper limit and (in brackets) 
99.9\% upper limit. All of the values are obtained from MCMC. 
The columns compare different theoretical priors and different data sets.
The parameters for 6 parameter models 6-p are
$\tau,\omega_b,\omega_m,\Omega_m=1-\Omega_{\lambda}, \sigma_8, n_s$.
\begin{center}
\begin{tabular}{|l|c|c|c|c|c|}
\hline 
&6-p & 6-p  &6-p   & 6-p+$r$& 6-p+$r$    \\
& & & & &   \\
\hline
&WMAP+gal& WMAP+gal+lya & all& WMAP+gal+lya & all   \\
& & & & &   \\
\hline
& & & & &   \\
$10^2\omega_b$ & 
$2.38^{+0.14}_{-0.12}\;{}^{+0.27}_{-0.23}\;{}^{+0.39}_{-0.33}$ &
$2.31^{+0.09}_{-0.08}\;{}^{+0.17}_{-0.17}\;{}^{+0.26}_{-0.24}$ &
$2.33^{+0.09}_{-0.08}\;{}^{+0.17}_{-0.17}\;{}^{+0.26}_{-0.25}$ &
$2.40^{+0.12}_{-0.105}\;{}^{+0.26}_{-0.19}\;{}^{+0.47}_{-0.30}$ &
$2.40^{+0.11}_{-0.10}\;{}^{+0.23}_{-0.19}\;{}^{+0.33}_{-0.27}$
\\
 & & & & &  \\
$\Omega_m$ & 
$0.294^{+0.041}_{-0.034}\;{}^{+0.089}_{-0.061}\;{}^{+0.143}_{-0.082}$ &
$0.299^{+0.037}_{-0.032}\;{}^{+0.082}_{-0.061}\;{}^{+0.133}_{-0.084}$ &
$0.281^{+0.023}_{-0.021}\;{}^{+0.046}_{-0.040}\;{}^{+0.070}_{-0.061}$ &
$0.278^{+0.036}_{-0.033}\;{}^{+0.076}_{-0.062}\;{}^{+0.118}_{-0.094}$ &
$0.270^{+0.022}_{-0.021}\;{}^{+0.045}_{-0.041}\;{}^{+0.072}_{-0.060}$

\\
 & & & & & \\
$n_s$ & 
$0.994^{+0.044}_{-0.035}\;{}^{+0.077}_{-0.060}\;{}^{+0.101}_{-0.080}$ &
$0.971^{+0.023}_{-0.019}\;{}^{+0.048}_{-0.038}\;{}^{+0.070}_{-0.055}$ &
$0.980^{+0.020}_{-0.019}\;{}^{+0.041}_{-0.037}\;{}^{+0.065}_{-0.051}$ &
$1.00^{+0.034}_{-0.028}\;{}^{+0.070}_{-0.050}\;{}^{+0.124}_{-0.076}$ &
$1.00^{+0.027}_{-0.024}\;{}^{+0.056}_{-0.045}\;{}^{+0.085}_{-0.063}$
\\
 & & & & & \\
$\tau$ & 
$0.176^{+0.078}_{-0.071}\;{}^{+0.117}_{-0.124}\;{}^{+0.124}_{-0.161}$ &
$0.133^{+0.052}_{-0.045}\;{}^{+0.104}_{-0.087}\;{}^{+0.148}_{-0.126}$ &
$0.160^{+0.040}_{-0.041}\;{}^{+0.079}_{-0.080}\;{}^{+0.117}_{-0.120}$ &
$0.138^{+0.050}_{-0.045}\;{}^{+0.096}_{-0.085}\;{}^{+0.151}_{-0.118}$ &
$0.155^{+0.040}_{-0.040}\;{}^{+0.078}_{-0.077}\;{}^{+0.112}_{-0.114}$

\\
 & & & & & \\
$\sigma_8$ & 
$0.951^{+0.090}_{-0.079}\;{}^{+0.173}_{-0.142}\;{}^{+0.124}_{-0.161}$ &
$0.890^{+0.034}_{-0.032}\;{}^{+0.065}_{-0.060}\;{}^{+0.096}_{-0.089}$ &
$0.897^{+0.033}_{-0.031}\;{}^{+0.065}_{-0.058}\;{}^{+0.097}_{-0.086}$ &
$0.901^{+0.035}_{-0.033}\;{}^{+0.069}_{-0.062}\;{}^{+0.107}_{-0.096}$ &
$0.904^{+0.035}_{-0.031}\;{}^{+0.069}_{-0.059}\;{}^{+0.106}_{-0.094}$
\\
 & & & & & \\
$h$ & 
$0.706^{+0.037}_{-0.034}\;{}^{+0.068}_{-0.065}\;{}^{+0.097}_{-0.091}$ &
$0.694^{+0.030}_{-0.028}\;{}^{+0.059}_{-0.057}\;{}^{+0.092}_{-0.086}$ &
$0.710^{+0.021}_{-0.021}\;{}^{+0.044}_{-0.040}\;{}^{+0.066}_{-0.061}$ &
$0.719^{+0.036}_{-0.032}\;{}^{+0.076}_{-0.061}\;{}^{+0.133}_{-0.091}$ &
$0.726^{+0.025}_{-0.023}\;{}^{+0.052}_{-0.045}\;{}^{+0.081}_{-0.068}$

\\
 & & & & & \\
$r$ & 0 & 0 & 0 & $<0.38 (0.55)$  & $<0.36 (0.51) $  
\\
 & & & & & \\

\hline
\end{tabular}
\end{center}
}
\label{table1}
\end{table*}

\begin{table*}
\noindent
{\footnotesize
Table 2: Constraints on running. 
Same format as for table 1.
\begin{center}
\begin{tabular}{|l|c|c|c|c|c|}
\hline 
& 6-p+$\alpha_s$  &6-p+$\alpha_s$   & 6-p+$\alpha_s$ & 6-p+$\alpha_s$  & 6-p+$\alpha_s + r$  \\
& & & & &   \\
\hline
& WMAP & WMAP+gal & WMAP+lya & all & WMAP+gal+lya \\
& & & & &   \\
\hline
& & & & &   \\
$10^2\omega_b$ & 
$2.33^{+0.16}_{-0.16}\;{}^{+0.33}_{-0.32}\;{}^{+0.50}_{-0.47}$ &
$2.30^{+0.14}_{-0.14}\;{}^{+0.29}_{-0.27}\;{}^{+0.45}_{-0.38}$ &
$2.36^{+0.11}_{-0.10}\;{}^{+0.22}_{-0.19}\;{}^{+0.32}_{-0.27}$ &
$2.33^{+0.09}_{-0.09}\;{}^{+0.18}_{-0.17}\;{}^{+0.28}_{-0.25}$&
$2.42^{+0.12}_{-0.12}\;{}^{+0.24}_{-0.22}\;{}^{+0.39}_{-0.31}$

\\
& & & & &   \\
$\Omega_m$ & 
$0.246^{+0.072}_{-0.057}\;{}^{+0.159}_{-0.103}\;{}^{+0.263}_{-0.140}$&
$0.269^{+0.041}_{-0.033}\;{}^{+0.091}_{-0.062}\;{}^{+0.156}_{-0.095}$ &
$0.257^{+0.055}_{-0.048}\;{}^{+0.105}_{-0.073}\;{}^{+0.151}_{-0.092}$&
$0.281^{+0.022}_{-0.021}\;{}^{+0.045}_{-0.043}\;{}^{+0.067}_{-0.062}$ &
$0.273^{+0.037}_{-0.033}\;{}^{+0.077}_{-0.059}\;{}^{+0.119}_{-0.089}$ 

\\
& & & & &  \\
$n_s$ & 
$0.977^{+0.061}_{-0.061}\;{}^{+0.122}_{-0.123}\;{}^{+0.181}_{-0.190}$&
$0.959^{+0.052}_{-0.053}\;{}^{+0.104}_{-0.107}\;{}^{+0.164}_{-0.161}$ &
$0.990^{+0.032}_{-0.029}\;{}^{+0.063}_{-0.053}\;{}^{+0.090}_{-0.076}$&
$0.977^{+0.025}_{-0.021}\;{}^{+0.052}_{-0.040}\;{}^{+0.083}_{-0.058}$ &
$1.00^{+0.034}_{-0.032}\;{}^{+0.070}_{-0.060}\;{}^{+0.102}_{-0.085}$ 
\\
& & & & &  \\
$\tau$ & 
$0.204^{+0.070}_{-0.086}\;{}^{+0.092}_{-0.149}\;{}^{+0.0957}_{-0.192}$&
$0.195^{+0.065}_{-0.068}\;{}^{+0.097}_{-0.123}\;{}^{+0.103}_{-0.165}$ &
$0.188^{+0.078}_{-0.075}\;{}^{+0.108}_{-0.130}\;{}^{+0.111}_{-0.171}$ &
$0.163^{+0.041}_{-0.041}\;{}^{+0.083}_{-0.078}\;{}^{+0.123}_{-0.111}$&
$0.142^{+0.0493}_{-0.0465}\;{}^{+0.0979}_{-0.0879}\;{}^{+0.143}_{-0.117}$

\\
& & & & &  \\
$\sigma_8$ & 
$0.873^{+0.115}_{-0.107}\;{}^{+0.24}_{-0.201}\;{}^{+0.381}_{-0.297}$ &
$0.897^{+0.059}_{-0.059}\;{}^{+0.108}_{-0.104}\;{}^{+0.189}_{-0.137}$ &
$0.895^{+0.034}_{-0.032}\;{}^{+0.068}_{-0.064}\;{}^{+0.102}_{-0.094}$ &
$0.899^{+0.034}_{-0.030}\;{}^{+0.070}_{-0.058}\;{}^{+0.107}_{-0.085}$&
$0.900^{+0.034}_{-0.032}\;{}^{+0.069}_{-0.063}\;{}^{+0.100}_{-0.094}$
\\
& & & & &  \\
$h$ & 
$0.736^{+0.061}_{-0.054}\;{}^{+0.127}_{-0.103}\;{}^{+0.204}_{-0.146}$&
$0.716^{+0.039}_{-0.040}\;{}^{+0.079}_{-0.080}\;{}^{+0.135}_{-0.121}$ &
$0.730^{+0.053}_{-0.046}\;{}^{+0.092}_{-0.080}\;{}^{+0.128}_{-0.107}$ &
$0.709^{+0.022}_{-0.021}\;{}^{+0.046}_{-0.040}\;{}^{+0.072}_{-0.059}$&
$0.725^{+0.037}_{-0.035}\;{}^{+0.074}_{-0.066}\;{}^{+0.123}_{-0.094}$

\\
& & & & &  \\
$r$ & 0 & 0 & 0 & 0 & $<0.45 (0.64)$ 
\\
& & & & &  \\
$10^2\alpha_s$ & 
$-1.24^{+3.75}_{-3.63}\;{}^{+7.63}_{-7.23}\;{}^{+11.8}_{-11.1}$ &
$-2.41^{+3.07}_{-3.10}\;{}^{+6.24}_{-6.14}\;{}^{+9.45}_{-9.20}$ &
$-0.263^{+1.27}_{-1.13}\;{}^{+2.66}_{-2.21}\;{}^{+4.15}_{-3.22}$ &
$-0.29^{+1.08}_{-1.00}\;{}^{+2.35}_{-1.84}\;{}^{+3.63}_{-2.61}$ &
$-0.57^{+1.21}_{-1.14}\;{}^{+2.49}_{-2.26}\;{}^{+3.48}_{-3.39}$

\\
& & & & & \\

\hline
\end{tabular}
\end{center}
}
\label{table2}
\end{table*}

\begin{table*}
\noindent
{\footnotesize
Table 3: Neutrino mass constraints. Same format as for table 1. 
All except last column are for the case of 3 degenerate neutrino families. 
Last column is for 3 massless + 1 massive neutrino family. 
\begin{center}
\begin{tabular}{|l|c|c|c|c|c|}
\hline 
& 6-p+$3\times m_{\nu}$  &6-p+$3\times m_{\nu}$  & 6-p+$3\times m_{\nu}$ & 6-p+$3\times m_{\nu}+\alpha_s+r$  & 6-p+1$\times m_{\nu}$ \\
& & & & &   \\
\hline
& WMAP+gal & WMAP+gal+lya & all &  all & all \\
& & & & &   \\
\hline
& & & & &   \\
$10^2\omega_b$ & 
$2.41^{+0.16}_{-0.14}\;{}^{+0.31}_{-0.25}\;{}^{+0.46}_{-0.37}$ &
$2.34^{+0.08}_{-0.08}\;{}^{+0.17}_{-0.17}\;{}^{+0.25}_{-0.25}$&
$2.36^{+0.09}_{-0.09}\;{}^{+0.19}_{-0.18}\;{}^{+0.32}_{-0.29}$&
$2.47^{+0.13}_{-0.12}\;{}^{+0.26}_{-0.22}\;{}^{+0.38}_{-0.32}$&
$2.35^{+0.12}_{-0.10}\;{}^{+0.25}_{-0.19}\;{}^{+0.36}_{-0.28}$
\\
& & & & &   \\
$\Omega_m$ & 
$0.352^{+0.131}_{-0.080}\;{}^{+0.241}_{-0.120}\;{}^{+0.334}_{-0.149}$&
$0.316^{+0.029}_{-0.027}\;{}^{+0.067}_{-0.052}\;{}^{+0.124}_{-0.080}$&
$0.284^{+0.025}_{-0.023}\;{}^{+0.05}_{-0.044}\;{}^{+0.079}_{-0.060}$&
$0.277^{+0.025}_{-0.023}\;{}^{+0.051}_{-0.045}\;{}^{+0.086}_{-0.064}$&
$0.287^{+0.028}_{-0.025}\;{}^{+0.060}_{-0.048}\;{}^{+0.103}_{-0.069}$

\\
& & & & &  \\
$n_s$ & 
$1.00^{+0.051}_{-0.041}\;{}^{+0.098}_{-0.071}\;{}^{+0.131}_{-0.095}$&
$0.978^{+0.023}_{-0.020}\;{}^{+0.051}_{-0.039}\;{}^{+0.069}_{-0.055}$&
$0.989^{+0.026}_{-0.023}\;{}^{+0.053}_{-0.042}\;{}^{+0.076}_{-0.060}$&
$1.020^{+0.033}_{-0.033}\;{}^{+0.066}_{-0.061}\;{}^{+0.094}_{-0.082}$&
$1.00^{+0.032}_{-0.025}\;{}^{+0.061}_{-0.047}\;{}^{+0.083}_{-0.067}$
\\
& & & & &  \\
$\tau$ & 
$0.133^{+0.081}_{-0.060}\;{}^{+0.144}_{-0.101}\;{}^{+0.165}_{-0.128}$&
$0.153^{+0.055}_{-0.042}\;{}^{+0.107}_{-0.075}\;{}^{+0.140}_{-0.101}$ &
$0.185^{+0.052}_{-0.046}\;{}^{+0.099}_{-0.089}\;{}^{+0.114}_{-0.125}$ &
$0.206^{+0.059}_{-0.058}\;{}^{+0.088}_{-0.105}\;{}^{+0.093}_{-0.143}$&
$0.195^{+0.059}_{-0.055}\;{}^{+0.096}_{-0.102}\;{}^{+0.104}_{-0.147}$

\\
& & & & &  \\
$\sigma_8$ & 
$0.786^{+0.119}_{-0.100}\;{}^{+0.230}_{-0.172}\;{}^{+0.301}_{-0.230}$ &
$0.873^{+0.035}_{-0.032}\;{}^{+0.066}_{-0.065}\;{}^{+0.099}_{-0.093}$ &
$0.890^{+0.035}_{-0.033}\;{}^{+0.071}_{-0.064}\;{}^{+0.098}_{-0.092}$&
$0.882^{+0.032}_{-0.030}\;{}^{+0.069}_{-0.057}\;{}^{+0.107}_{-0.087}$&
$0.895^{+0.035}_{-0.033}\;{}^{+0.067}_{-0.063}\;{}^{+0.10}_{-0.094}$

\\
& & & & &  \\
$h$ & 
$0.663^{+0.070}_{-0.076}\;{}^{+0.117}_{-0.113}\;{}^{+0.164}_{-0.146}$&
$0.684^{+0.023}_{-0.022}\;{}^{+0.047}_{-0.047}\;{}^{+0.070}_{-0.083}$&
$0.710^{+0.023}_{-0.022}\;{}^{+0.047}_{-0.044}\;{}^{+0.075}_{-0.067}$&
$0.723^{+0.027}_{-0.025}\;{}^{+0.054}_{-0.047}\;{}^{+0.082}_{-0.080}$ &
$0.744^{+0.024}_{-0.023}\;{}^{+0.050}_{-0.047}\;{}^{+0.078}_{-0.072}$

\\
& & & & &  \\
$r$ & 0 & 0 & 0 & $<0.47 (0.63)$ & 0 
\\
& & & & &  \\
$10^2\alpha_s$ & 
0&0&0&$-0.18^{+1.23}_{-1.24}\;{}^{+2.46}_{-2.50}\;{}^{+3.78}_{-3.62}$& 0

\\
& & & & & \\
$\sum m_{\nu}$ & 
1.54 (2.26) eV & 0.54 (0.86) eV& 0.42 (0.67) eV &0.66 (0.93) eV & 0.84(1.61) eV \\
& & & & & \\

\hline
\end{tabular}
\end{center}
}
\label{table3}
\end{table*}

\begin{table*}
\noindent
{\footnotesize
Table 4: Dark energy constraints. Same format as for table 1. 
All columns except last one assume constant equation of state w. 
Last column gives constraints for the case where dark energy is 
time dependent as ${\rm w}={\rm w}_0+{\rm w}_1(1-a)$. 
\begin{center}
\begin{tabular}{|l|c|c|c|c|c|}
\hline 
& 6-p+w  &6-p+w  & 6-p+w & 6-p+w+$\alpha_s+r$  & 6-p+${\rm w}_0+{\rm w}_1$ \\
& & & & &   \\
\hline
& WMAP+gal+SN &all & WMAP+gal+bias+lya &WMAP+gal+bias+lya & all \\
& & & & &   \\
\hline
& & & & &   \\
$10^2\omega_b$ & 
$2.36^{+0.13}_{-0.11}\;{}^{+0.26}_{-0.21}\;{}^{+0.38}_{-0.31}$&
$2.33^{+0.10}_{-0.09}\;{}^{+0.20}_{-0.18}\;{}^{+0.32}_{-0.27}$&
$2.34^{+0.09}_{-0.09}\;{}^{+0.19}_{-0.16}\;{}^{+0.28}_{-0.25}$ &
$2.48^{+0.15}_{-0.13}\;{}^{+0.29}_{-0.24}\;{}^{+0.43}_{-0.34}$ &
$2.33^{+0.10}_{-0.09}\;{}^{+0.20}_{-0.17}\;{}^{+0.32}_{-0.25}$
\\
& & & & &   \\
$\Omega_m$ & 
$0.303^{+0.029}_{-0.028}\;{}^{+0.061}_{-0.052}\;{}^{+0.093}_{-0.072}$&
$0.282^{+0.023}_{-0.023}\;{}^{+0.047}_{-0.044}\;{}^{+0.074}_{-0.067}$&
$0.264^{+0.028}_{-0.025}\;{}^{+0.056}_{-0.046}\;{}^{+0.109}_{-0.062}$&
$0.260^{+0.024}_{-0.022}\;{}^{+0.050}_{-0.040}\;{}^{+0.077}_{-0.056}$ &
$0.285^{+0.024}_{-0.023}\;{}^{+0.047}_{-0.045}\;{}^{+0.070}_{-0.066}$

\\
& & & & &  \\
$n_s$ & 
$0.987^{+0.041}_{-0.030}\;{}^{+0.077}_{-0.054}\;{}^{+0.105}_{-0.075}$ &
$ 0.981^{+0.027}_{-0.023}\;{}^{+0.055}_{-0.042}\;{}^{+0.080}_{-0.062}$&
$0.980^{+0.026}_{-0.020}\;{}^{+0.051}_{-0.038}\;{}^{+0.068}_{-0.059}$ &
$1.020^{+0.041}_{-0.037}\;{}^{+0.080}_{-0.068}\;{}^{+0.114}_{-0.096}$ &
$0.978^{+0.028}_{-0.022}\;{}^{+0.058}_{-0.041}\;{}^{+0.084}_{-0.059}$
\\
& & & & &  \\
$\tau$ & 
$ 0.160^{+0.082}_{-0.067}\;{}^{+0.130}_{-0.116}\;{}^{+0.139}_{-0.153}$&
$0.163^{+0.064}_{-0.057}\;{}^{+0.121}_{-0.103}\;{}^{+0.135}_{-0.146}$&
$0.145^{+0.066}_{-0.056}\;{}^{+0.125}_{-0.109}\;{}^{+0.152}_{-0.142}$ &
$0.201^{+0.057}_{-0.063}\;{}^{+0.091}_{-0.117}\;{}^{+0.098}_{-0.163}$ &
$0.152^{+0.067}_{-0.056}\;{}^{+0.127}_{-0.101}\;{}^{+0.146}_{-0.136}$

\\
& & & & &  \\
$\sigma_8$ & 
$ 0.945^{+0.089}_{-0.080}\;{}^{+0.187}_{-0.150}\;{}^{+0.290}_{-0.212} $&
$0.895^{+0.033}_{-0.031}\;{}^{+0.067}_{-0.059}\;{}^{+0.104}_{-0.089}$&
$0.920^{+0.040}_{-0.041}\;{}^{+0.084}_{-0.072}\;{}^{+0.12}_{-0.093}$ &
$0.890^{+0.030}_{-0.028}\;{}^{+0.063}_{-0.056}\;{}^{+0.099}_{-0.089}$ &
$0.897^{+0.033}_{-0.031}\;{}^{+0.068}_{-0.059}\;{}^{+0.104}_{-0.088}$

\\
& & & & &  \\
$h$ & 
$ 0.699^{+0.027}_{-0.026}\;{}^{+0.054}_{-0.050}\;{}^{+0.080}_{-0.073}$&
$0.708^{+0.023}_{-0.022}\;{}^{+0.046}_{-0.044}\;{}^{+0.069}_{-0.064}$&
$0.736^{+0.039}_{-0.038}\;{}^{+0.080}_{-0.069}\;{}^{+0.119}_{-0.112}$ &
$0.726^{+0.025}_{-0.024}\;{}^{+0.050}_{-0.048}\;{}^{+0.078}_{-0.072}$ &
$0.707^{+0.024}_{-0.023}\;{}^{+0.049}_{-0.046}\;{}^{+0.074}_{-0.066}$

\\
& & & & &  \\
$r$ & 0 & 0 & 0 & $<0.51 (0.67)$ & 0 
\\
& & & & &  \\
$10^2\alpha_s$ & 
0&0&0&$-1.07^{+1.24}_{-1.16}\;{}^{+2.64}_{-2.26}\;{}^{+4.15}_{-3.31}$&0
\\
& & & & & \\
w & 
$-1.009^{+0.096}_{-0.112}\;{}^{+0.18}_{-0.24}\;{}^{+0.26}_{-0.38}$&
$-0.990^{+0.086}_{-0.093}\;{}^{+0.16}_{-0.20}\;{}^{+0.22}_{-0.35}$&
$-1.080^{+0.149}_{-0.193}\;{}^{+0.24}_{-0.37}\;{}^{+0.31}_{-0.54}$& 
$-0.908^{+0.077}_{-0.091}\;{}^{+0.14}_{-0.19}\;{}^{+0.19}_{-0.32}$ &
$-0.981^{+0.193}_{-0.193}\;{}^{+0.38}_{-0.37}\;{}^{+0.57}_{-0.52}$
\\
& & & & & \\
${\rm w_1}$ &0&0&0&0&
$0.05^{+0.83}_{-0.65}\;{}^{+1.92}_{-1.13}\;{}^{+2.88}_{-1.38}$ 
\\
& & & & & \\
\hline
\end{tabular}
\end{center}
}
\label{table4}
\end{table*}

\subsection{Amplitude of fluctuations}

From tables 1-4 one can see that the value of $\sigma_8$ is 
remarkably tight. For 6-p models (table 1) we find 
\be
\sigma_8=0.897^{+0.033}_{-0.031}\;{}^{+0.065}_{-0.058}\;{}^{+0.097}_{-0.088} 
\ee
This value does not change significantly when running, tensors
and massive neutrinos are added to the mix, which shows that 
the constraint is model independent. 
In contrast, in 
an analysis without the \lyaf\ and bias $\sigma_8$ changes from 
$\sigma_8=0.951^{+0.90}_{-0.079}$ (table 1) to $\sigma_8=0.786^{+0.119}_{-0.100}$
(table 3) 
when massive neutrinos are added as a parameter (see also 
\cite{2004PhRvD..69j3501T}), 
so 
previous constraints were significantly more model dependent. 

It is useful to analyze what drives the $\sigma_8$ determination. 
WMAP alone cannot provide a very tight determination, nor can 
the \lyaf\ alone. But combining the two is extremely powerful: 
from table 1 we see that just these two data sets alone give 
$\sigma_8=0.895^{+0.034}_{-0.032}$ even with running. 
So this combination in itself provides nearly all of the 
information on $\sigma_8$; galaxy clustering and bias do not 
constrain this parameter any further when added to the mix. 
They are however consistent with it: 
using WMAP and SDSS galaxy clustering 
with bias and {\it without} \lyaf\ gives 
$\sigma_8=0.89\pm 0.06$ \cite{2004astro.ph..6594S}, 
in remarkable agreement with the analysis of WMAP+SDSS-lya.
Assuming that WMAP data are valid 
this implies that two independent analyses of different data, 
SDSS-gal+bias and SDSS-lya, lead to essentially 
the same value. Both improve upon 
previous constraints,
by a factor of 1.5-2 for WMAP+SDSS-gal+bias and a factor of 3-4 for WMAP+SDSS-lya. 
These new constraints remove almost all of the degeneracy between $\sigma_8$ 
and optical depth $\tau$ (figure \ref{fig5}). 

There are many recent determinations of $\sigma_8$ in the literature, 
which vary between 0.6 and 1.1. 
Recent discussion of some of these methods and results, such as weak lensing, 
cluster abundance, galaxy bias determination, and SZ power spectrum 
can be found in \cite{2004PhRvD..69j3501T,2004astro.ph..6594S}. 
The value found here is in good 
agreement with most of these constraints: it is on the low 
end of the SZ constraints and on the upper end of some of the 
cluster abundance constraints. It is also in good agreement with the
2dF bias constraints and with several weak lensing constraints. 

While in the tables we do not present results for the amplitude of metric 
(described here with curvature fluctuation ${\cal R}$) 
fluctuations at the pivot point we find it 
is also tightly constrained to 
\be
\Delta_{\cal R}^2(k_{\rm pivot}=0.05/{\rm Mpc})=(2.45\pm 0.23)\times 10^{-9}.
\ee 

\subsection{Optical depth}

The optical depth due to reionization is a parameter that has a strong 
effect on the CMB. It suppresses the CMB on small scales and thus leads 
to a strong degeneracy with amplitude. 
This degeneracy can be lifted by the polarization observations 
\cite{1997PhRvD..55.1822Z}, but for WMAP 1st year these are noisy and may contain significant 
contamination from foregrounds. The current analysis based on 1st year data 
is rather unsatisfactory, since it is based on the existing
temperature-polarization 
cross-correlation analysis, which on large scales 
may suffer from similar 
problems as the temperature auto-correlation 
analysis \cite{2004astro.ph..3073S}. 
The upcoming 2nd year data 
release of WMAP should provide polarization maps and the 
corresponding analysis may help improve the situation. 
Until then we will use the current WMAP provided likelihood code \cite{2003ApJS..148..195V}, 
but this should be taken as preliminary and the constraints on
optical depth
from polarization, both the best fitted value and the associated errors, 
may change. 

\begin{figure}
\noindent
\includegraphics[width=.48\textwidth]{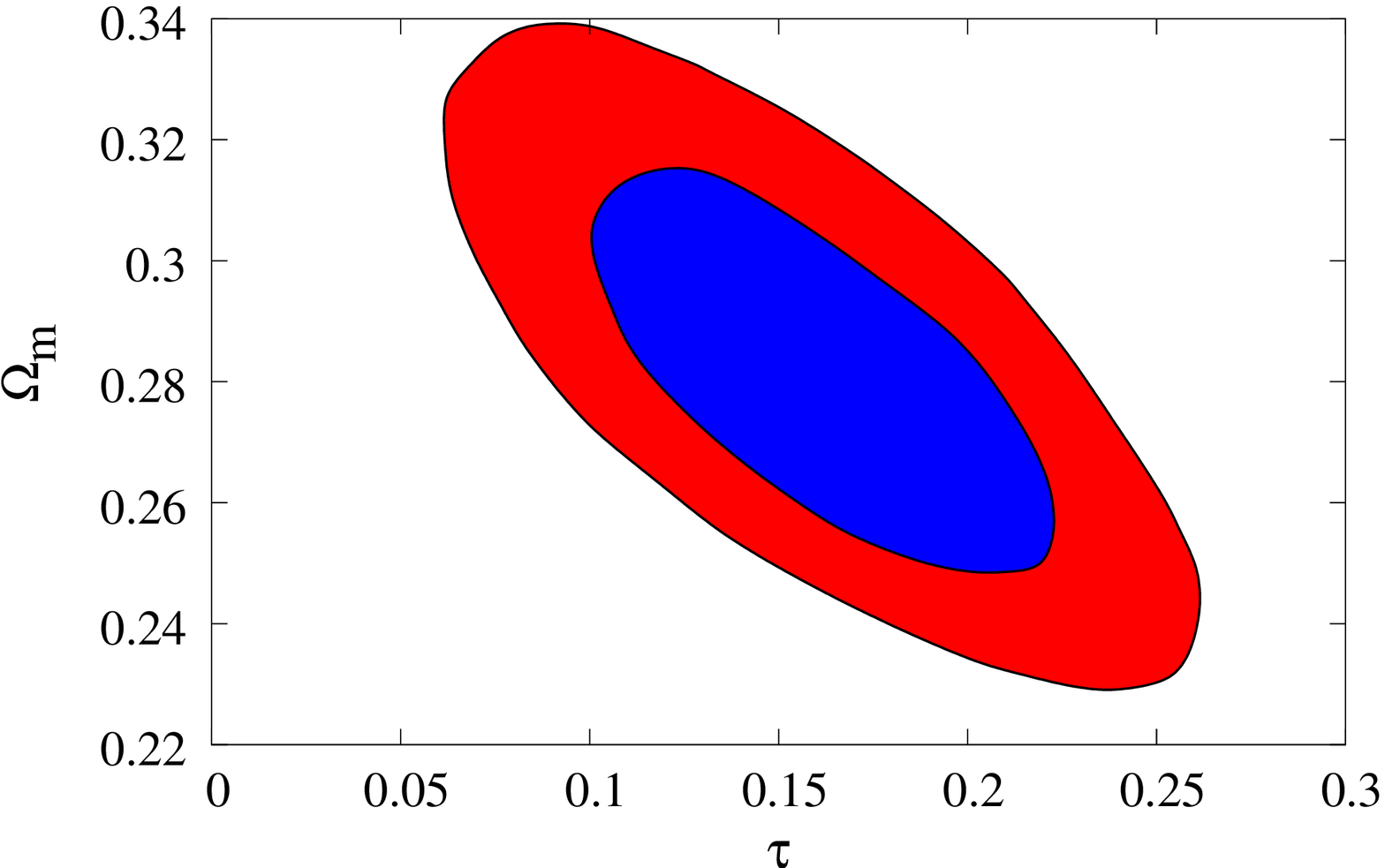}\hfill%
\includegraphics[width=.48\textwidth]{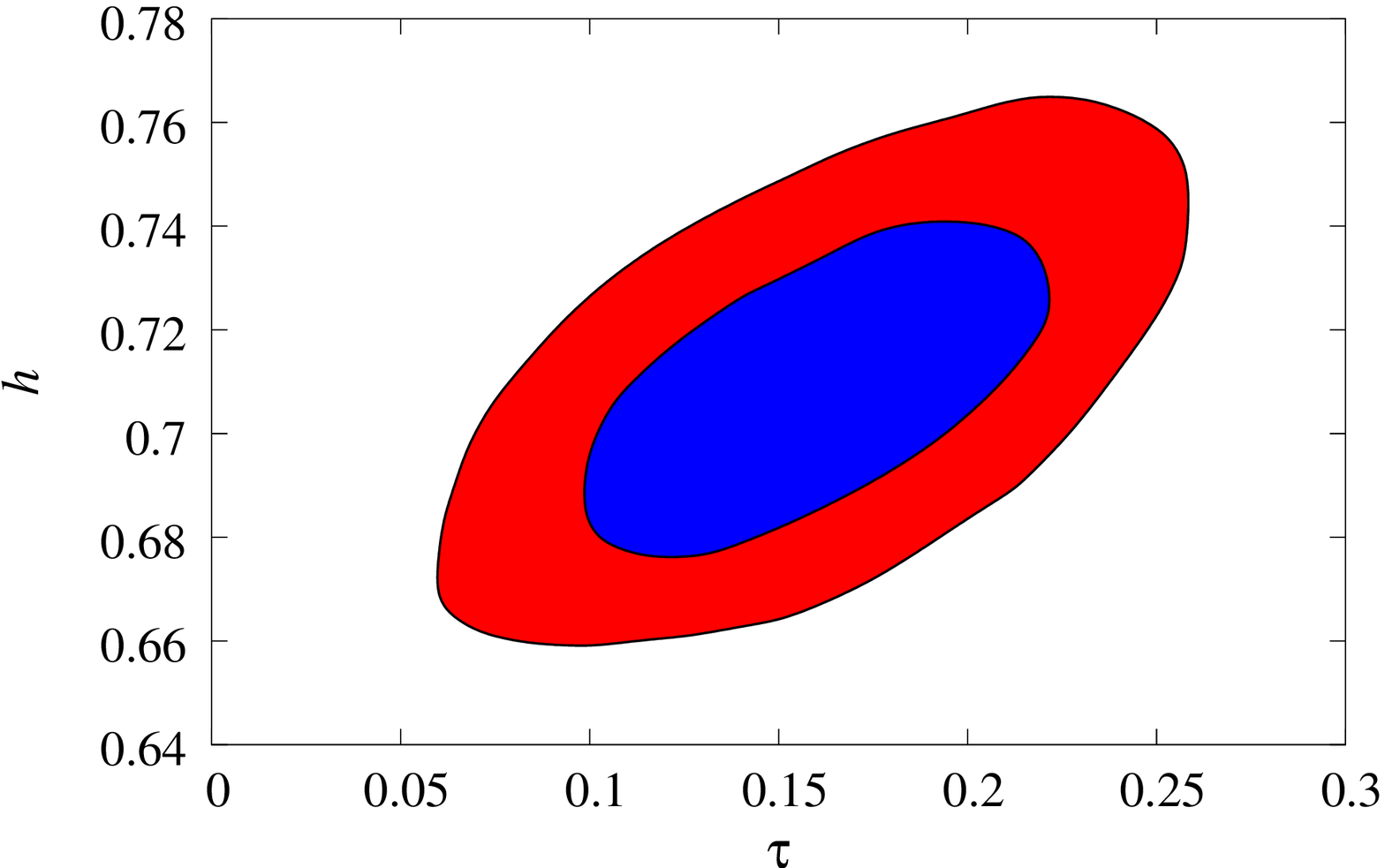}\\[5mm]
\includegraphics[width=.48\textwidth]{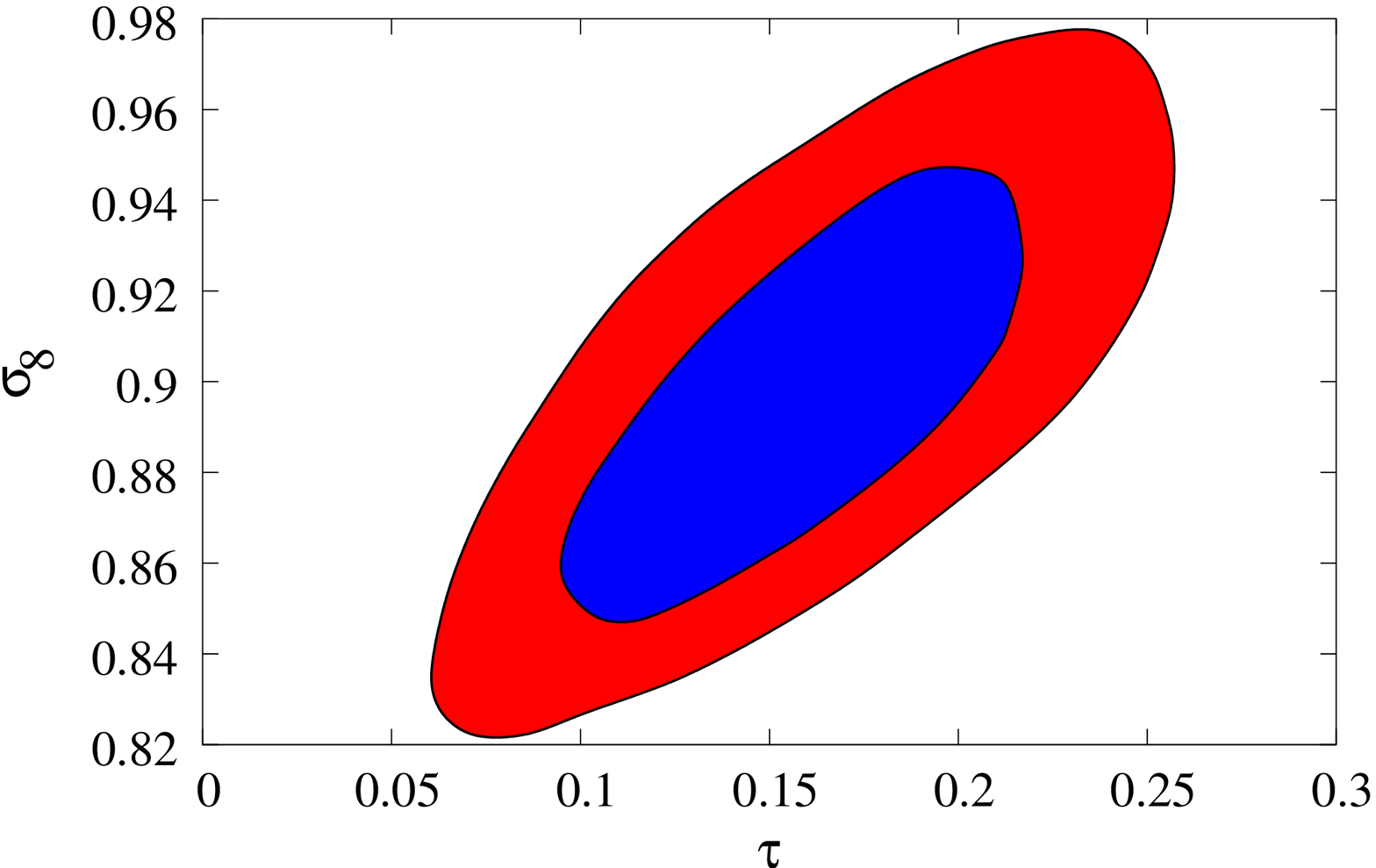}\hfill%
\includegraphics[width=.48\textwidth]{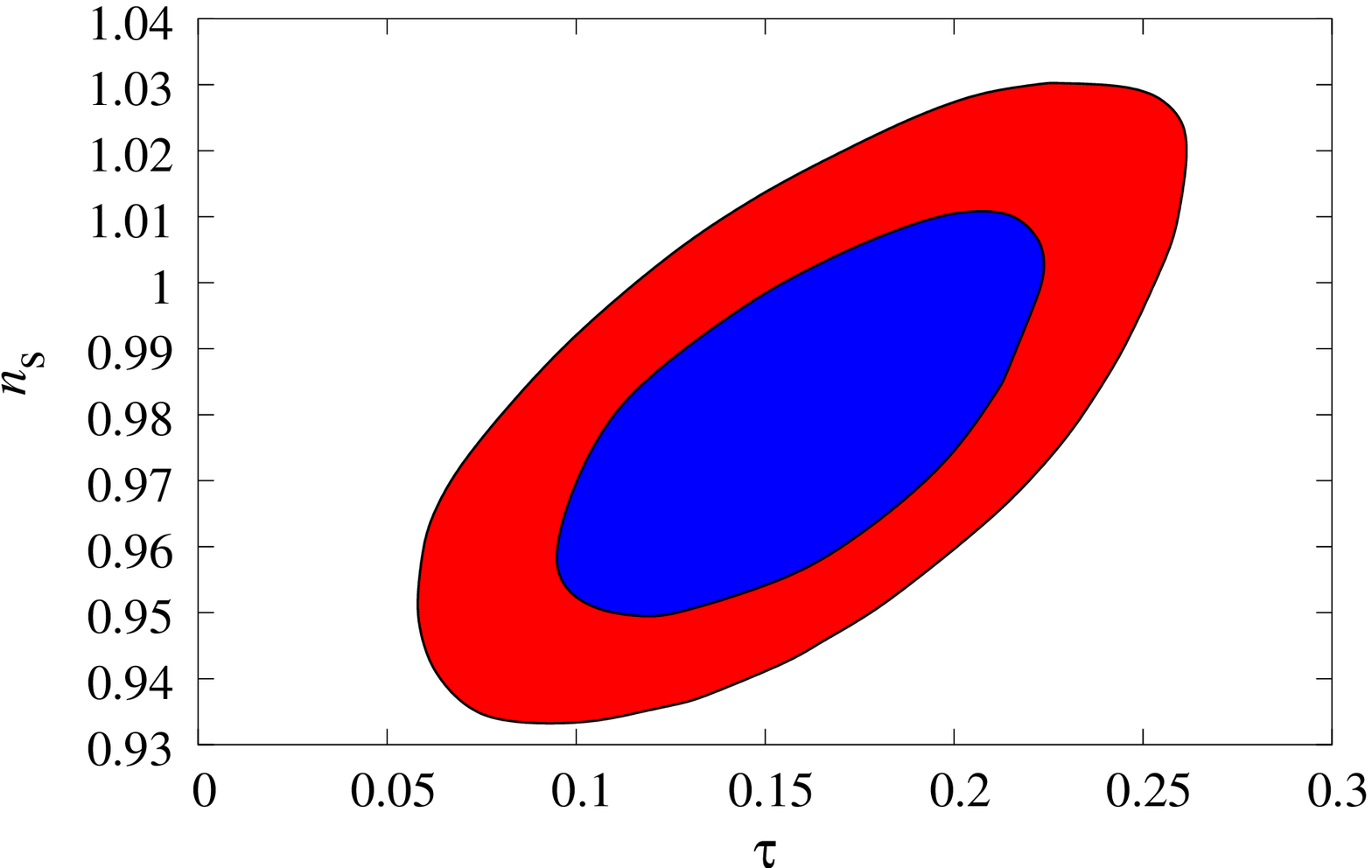}

\caption{68\% (inner, blue) and 95\% (outer, red) 
contours in the plane of $\tau)$ versus $\Omega_m$, $h$, $\sigma_8$
and $n_s$, respectively,
using all measurements. 
}
\label{fig5}
\end{figure}

With the addition of new constraints from the \lyaf\ and SDSS bias
there remain
correlations between optical depth $\tau$ and several other 
parameters from 6-parameter analysis on all data in table 1. Results are 
shown in figure \ref{fig5}. The degeneracies are 
significantly less severe than before,
since the parameters are better determined with the new data.  
Still, there is room to improve the constraints 
with a better determination of the optical depth. For example, if 
the optical depth ends up being at the lower end of its allowed range this 
would lead to a decrease in the best fitted value of $n_s$, $h$ and $\sigma_8$ 
and to an increase in the best fitted value of $\Omega_m$. 
Note that the values of $\tau$ do not extend up to the cutoff value $\tau=0.3$ 
for the 95\% contours, so these
distributions are not affected by the choice of the prior $\tau<0.3$. 
However, in chains with more parameters, such as dark energy 
equation of state w, this is no longer the case. 
At the moment the only argument for adopting this prior is that if
$\tau>0.3$ this would possibly have led to detectable auto-correlation 
of polarization in the WMAP data, but this argument is inconclusive since 
the polarization maps are not available and such analysis has not been 
published yet. In the absence of any published 
results we follow the WMAP team approach and adopt $\tau<0.3$.  

\subsection{Neutrino mass} 

Both the CMB and LSS 
are important as tracers of neutrino mass.  
At the time of decoupling, 
neutrinos are still relativistic,
but
become nonrelativistic later in the evolution of the universe if 
their mass is sufficiently high.
Neutrinos free-stream out of their 
potential wells, erasing their own perturbations on smaller scales. 
Below this suppression scale the power spectrum shape is the same as in 
regular CDM models, so on small scales 
the only consequence is the suppression of the amplitude relative 
to large scales. 
In the matter power spectrum neutrinos leave a characteristic 
feature at the transition scale.
The actual shape of the transition depends on the 
individual masses of neutrinos and not just on their sum. 
For masses of interest today the transition is occurring 
around $k=0.1h$/Mpc, which are the scales measured by SDSS-gal. 
Neutrinos with mass below 2eV 
are still relativistic when they enter the horizon for 
scales around $k=0.1 h$/Mpc and are either relativistic or 
quasi-relativistic at the time of recombination, $z\sim 1100$. 
As a result neutrinos cannot 
be treated as a nonrelativistic component with regard to the CMB and 
are not completely degenerate with the other relativistic components in
the CMB.

From the joint analysis we find for the sum of all masses (table 3)
\be
\sum m_{\nu}<0.42{\rm eV}~(0.67{\rm eV}) ~{\rm (3~families)},
\label{mnu}
\ee
at 95\% (99.9\%) c.l. for a 
single component and assuming no running, as was
done in all of the work to date. 
Our constraints improve upon WMAP+SDSS-gal, where 
we find 
$m_{\nu}<1.54$eV and upon 
WMAP+2dF
constraints, where $m_{\nu}<0.69$eV was found
by combining WMAP and 2dF with the
bias determination from the 
bispectrum analysis \citep{2002MNRAS.335..432V}. 

If running and tensors are allowed, the parameter space expands. In 
this case, we find 
$m_{\nu}<0.66~(0.93){\rm eV}$.
Much of this is caused by running: 
as discussed in \cite{2004astro.ph..6594S} running and neutrino mass
are anti-correlated. 
Negative runnings as large as -0.04 and neutrino masses as high as 1.5eV 
are allowed at 2-sigma. 
Running is poorly motivated by inflationary models and there is 
no evidence for it in the current data, so adopting the inflationary
prior with no running is reasonable, but one should be aware 
that the limits are model dependent. 

The constraint from equation \ref{mnu} is remarkably tight and 
implies the upper limit on neutrino mass assuming degeneracy is 0.14eV
at 95\% c.l. 
Our constraint 
has been obtained assuming 3 degenerate mass neutrino families, 
but if the neutrino mass splittings are small the constraints on 
the sum are almost the same even if individual masses are not identical. 
If the masses are very large compared to mass 
splittings then the neutrino masses are close to degenerate. 
However, our upper limit is so low that including mass splittings 
is necessary.
Super-Kamionkande (SK) results find neutrino
mass squared difference 
$\delta m_{23}=2.5\times 10^{-3}{\rm eV}^2$ 
\cite{1998PhRvL..81.1158F,2004hep.ex..4034A}, 
while solar neutrino constraints find neutrino
mass squared difference $\delta m_{12}=8\times 10^{-5}{\rm eV}^2$ 
\cite{2002PhRvC..66c5802B,2001PhRvL..87g1301A,2004hep.ph..6294B}. 
This gives one neutrino family with minimum mass around 0.05eV 
and another with minimum mass close to 0.007eV.
Since only the mass difference is measured,
it is in principle possible that the actual neutrino masses
are larger than that. 
Our constraints in combination with SK and solar neutrino constraints 
limit the mass of 
the neutrino families to
\be
m_1<0.13 {\rm eV}, \,\,\,\,\, m_2<0.13 {\rm eV},\,\,\,\,\, m_3<0.14 {\rm eV}, 
\ee
all at 95\% c.l.
These limits essentially exclude the range of masses argued by 
the Heidelberg-Moscow experiment of neutrinoless double beta decay 
if neutrinos are Majorana particles
\cite{2003idm..conf..577K}, although the two results may still 
be compatible given all the uncertainties in nuclear matrix element 
calculations.
From $\Delta m/m\sim \Delta m^2/2m$ with $m \sim 0.13$eV
we find the neutrino masses are not degenerate, but the limits 
are still weak: 
the ratios must satisfy 
\be
{m_3 \over m_1}>1.1,\,\,\,\,\, 1.1<{m_3 \over m_2}<7, 
\ee 
where the upper limit on $m_3/m_2$ is determined solely from 
SK and solar neutrino constraints.  

The mass limits presented above are based on 3 degenerate 
massive neutrino families. If one assumes a model with 3 massless
families and 1 massive family (such as a sterile neutrino model),
as motivated by LSND results \cite{1996PhRvL..77.3082A},
then the mass limits on the sum change, since both the CMB 
and the matter power spectrum change (see figure 6 
in \cite{2004astro.ph..6594S}). These 
limits are improved as well with the addition of SDSS-lya and SDSS-bias. 
We find 
\be
m_{\nu}<0.79{\rm eV}~(1.55{\rm eV}) ~{\rm (3+1~families)},
\label{sterile}
\ee
at 95 \% (99.9\%), 
compared to the WMAP+2dF analysis without bias where
the 95\% confidence 
limit is 1.4eV \cite{2003JCAP...05..004H} and to the SDSS+WMAP analysis
where the limit is 1.37eV \cite{2004astro.ph..6594S}.
We have subtracted from the total sum in table 3 
the masses of the active neutrinos to obtain the limit in equation \ref{sterile}. 
These limits are improved by almost a factor of 2 compared to previous 
analyses. 
These limits are more model independent,
as there is little correlation with running and/or 
tensors in 
this model: for the chains with running and tensors we find
$m_{\nu}<0.88{\rm eV}~(1.40{\rm eV})$ at 95\% (99.9\%) c.l. 

From the LSND experiment the allowed regions are four islands with the lowest mass 
$m_{\nu}=0.9$eV  and the next lowest 1.4eV \cite{2003NuPhS.114..203M,1996PhRvL..77.3082A,2003JCAP...05..004H,2004PhLB..581..218P}. 
Thus the lowest island allowed by 
LSND results is excluded at 95\% c.l. and all the others at 99.9\%. 
Our derived limits will be tested directly with MiniBoone 
Experiment at Fermilab \cite{2002NuPhS.110..420S}. 

\subsection{Tensors}

Gravity waves (tensors) are predicted in many models of inflation. 
The simplest single field models of inflation 
predict a tight relation between tensor amplitude and slope, which 
we assume here.  
We choose to parametrize them at the pivot point $k=0.05$/Mpc,
just as for the amplitude, slope
and running.
This pivot differs from that in the WMAP analysis \cite{2003ApJS..148..213P}.
While tensors have their largest effect on large scales, 
within the single field model adopted here the 
slope is assumed to be determined from the tensor amplitude. 
Thus there is no need to parametrize tensors on 
large scales. 

For 7-parameter model without running or neutrino mass, 
the limit on tensors is (table 1) 
\be
{T \over S}<0.36 (0.51)
\ee 
at 95\% (99.9\%) c.l. This does 
not change significantly if  
neutrinos or running are added to the mix (tables 2-3), in the latter 
case we find $r<0.45 (0.64)$. 
This constraint is nearly a factor of two better than from WMAP 
analysis, a consequence of tighter constraint on running from the \lyaf. 
We return to these constraints below where we discuss inflation. 

\subsection{Spectral index}  
Constraints on the scalar 
spectral index are primarily driven by the WMAP and SDSS-lya
combination. Using these two experiments alone one finds 
$n_s=0.990^{+0.032}_{-0.029}$ for the chains with running, compared to 
$n_s=0.962^{+0.054}_{-0.056}$ for WMAP+SDSS-gal+SDSS-bias 
without SDSS-lya and to $n_s=0.975^{+0.028}_{-0.024}$ for the case 
where all observations are included (table 2). The inclusion of the
SDSS \lyaf\ 
thus reduces the error on the primordial slope by a factor of 2. 
In the absence of running and with bias and SNIa, 
this constraint improves further to 
\be
n_s=0.981^{+0.019}_{-0.018}\;{}^{+0.040}_{-0.037}\;{}^{+0.061}_{-0.053}. 
\ee
Note that the scale invariant model $n_s=1$ is only 1-sigma away 
from the best fit.
It is remarkable that such a vast range of observational 
constraints can be reproduced with a scale invariant power spectrum 
with 4 parameters only, 
$\Omega_b$, $\Omega_m$, $h$ and amplitude $\Delta_{\cal R}^2$ (plus 
possibly optical depth $\tau$ to explain the polarization data). 

\begin{figure}
\centerline{\psfig{file=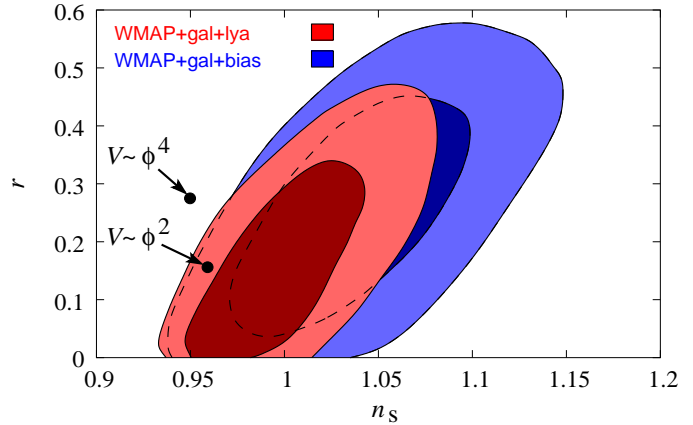,width=3.5in}}
\caption{68\% (inner, dark) and 95\% (outer, light)
contours in the $(r=T/S,n_s)$ plane
with and without SDSS-lya. 
There is a correlation between 
tensors and slope $n_s$. 
Inclusion of the \lyaf\ significantly reduces the allowed region in this plane. 
Also shown are the positions of 
two chaotic inflation 
models, $V\propto \phi^2$ with $N=50$ and $V \propto \phi^4$ with 
$N=60$. 
}
\label{fig6}
\end{figure}

Tensors are positively correlated 
with the slope (figure \ref{fig6}) and their inclusion
increases the best fit slope value to 
$n_s=1.00^{+0.034}_{-0.028}$.
All of these are consistent with a scale 
invariant spectrum and are in a good agreement with 
the WMAPext+2dF constraint $n_s=0.97\pm 0.03$ \cite{2003ApJS..148..175S}. 
While 2dF gives a slightly redder spectrum than SDSS 
the differences in different values quoted in the literature
reflect mostly the 
differences in the assumed parameter space, as shown here for 
the example of tensors.  

\subsection{Running of the spectral index}  

The issue of the running of the primordial slope has generated a 
lot of interest lately. WMAP argued for some weak evidence for 
negative running in their combined analysis, but some of that evidence 
was based on Lyman alpha constraints by previous workers
\citep{2002ApJ...581...20C,2002MNRAS.334..107G}, which were 
shown to underestimate the errors \citep{2003MNRAS.342L..79S}. 
It was argued that even from 
WMAP alone, or WMAP+2dF, there is some evidence for running,
and the WMAP+SDSS-gal analysis without bias information 
gave $\alpha_s=-0.071\pm 0.044$ \citep{2004PhRvD..69j3501T}. 
Similar values have been found from the recent analyses including 
CBI \cite{2004astro.ph..2359R} and VSA \cite{2004astro.ph..2466R} data. 
However, much of this effect comes from low $l$ multipoles and a full 
likelihood analysis of WMAP+SDSS-gal
changes this value to $-0.022\pm 0.033$ \cite{2004astro.ph..3073S}. 
Including the biasing constraints does not really change this 
result. In the absence of massive neutrinos and tensors 
we find $\alpha_s=-0.022^{+0.030}_{-0.032}$, so 
$\alpha_s=0$ is within one sigma of 0 and the error has not 
been reduced. 

\begin{figure}
\centerline{\psfig{file=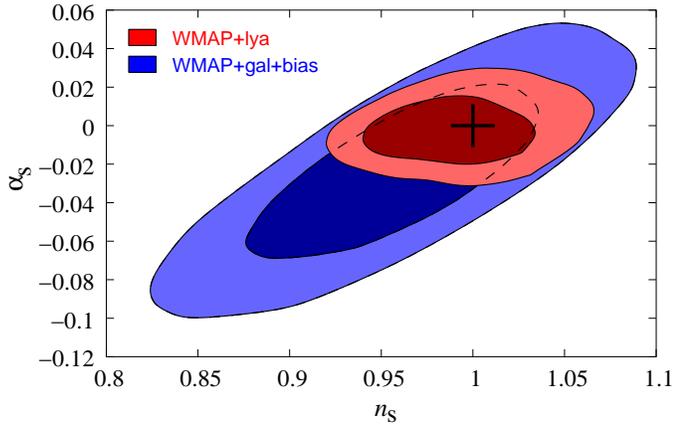,width=3.5in}}
\caption{68\% (inner, dark) and 95\% (outer, light)
contours in the $(\alpha_s,n_s)$ plane
using WMAP+SDSS-lya versus WMAP+SDSS-gal+bias. 
Adding the SDSS \lyaf\ dramatically reduces the allowed region
of parameter space in this plane. 
Note that the simplest model with $n_s=1$ and $\alpha_s=0$ is within 
68\% interval. 
}
\label{fig4}
\end{figure}

Including SDSS-lya reduces the errors dramatically. The 
constraint on running from WMAP+SDSS-lya alone is 
$\alpha_s=-0.0026^{+0.013}_{-0.011}$. Including everything 
this changes slightly to 
\be
\alpha_s=-0.0029^{+0.011}_{-0.010}\;{}^{+0.023}_{-0.018}\;{}^{+0.036}_{-0.026},
\ee
which is a factor of 3 improvement over previous constraints. 
Even with this significant improvement we find no hint of 
running in the joint analysis. 
The result is in perfect agreement 
with no running and 95\% of chain elements have $\alpha_s>-0.015$. 
This should be compared to values as low as $\alpha_s \sim -0.10$
in figure \ref{fig4}. Similarly low values have been 
found in recent analyses \cite{2004astro.ph..2359R,2004astro.ph..2466R}. 
Figure \ref{fig4} shows old and new
constraints in the $(\alpha_s,n_s)$ plane, 
highlighting the dramatic reduction of 
available parameter space when CMB and \lyaf\ data are combined together. 
The implications of this result for 
inflation are discussed in the next section. 

\begin{figure}
\centerline{\psfig{file=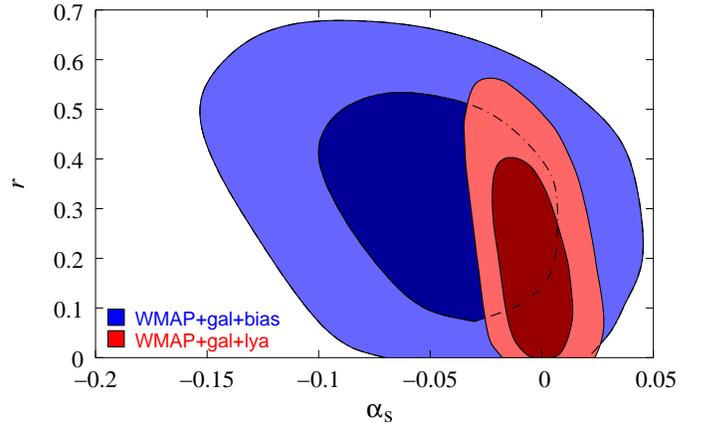,width=3.5in}}
\caption{68\% (inner, dark) and 95\% (outer, light)
contours in the $(\alpha_s,r)$ plane
using WMAP+SDSS-lya versus WMAP+SDSS-gal+bias. 
Adding the SDSS \lyaf\ dramatically reduces the allowed region
of parameter space in this plane. 
Note that the simplest model with $\alpha_s=0$ and $r=0$ is within 
the 68\% interval. 
}
\label{fig7}
\end{figure}

If tensors are also included 
they induce weak anti-correlation with running, so the best fit value becomes 
$\alpha_s=-0.006^{+0.012}_{-0.011}$, which is still perfectly consistent
with no running. This is shown in figure \ref{fig7}, where we see that 
adding SDSS-lya to the mix dramatically reduces the allowed 
region of parameter space. Specifically, without SDSS-lya, runnings as 
negative as -0.15 are in the 95\% confidence region, a consequence of 
strong correlation between running and tensors. 
Our joint analysis eliminates these large negative running solutions. 
We find no evidence 
for running in the current data, with or without tensors,
despite a factor of 3 reduction in the errors. 

Running is correlated with some of the "nuisance" parameters we 
marginalize over in the analysis and additional
observations constraining these could lead to a further reduction
of errors on the primordial slope and its running
even with no additional improvements in the observations.
For example, in our current treatment of the filtering parameter $k_F$
(a generalization of the Jeans length), we assume that the minimum 
reionization redshift is around 10 with a reheating temperature of 
25,000K. If we change the redshift to 7, this leads to an 
increase in the maximum value of $k_F$ allowed. In this case we find 
for WMAP+SDSS-lya analysis the running changes from $\alpha_s=0.0017$
to -0.0045, with an error around 0.01 (see table 2). 
If we change this redshift to 4, below its theoretically allowed 
lower limit of 6.5, to allow for any residual 
resolution issues in numerical simulations, we find $\alpha_s=-0.009$
with comparable errors. All the other parameters change much less. 
While these changes 
are small and do not qualitatively change our conclusions, 
they may be important for the future analyses where smaller 
errors may be obtained. In all these 
cases the data prefer a high value of $k_F$, 
i.e. a late epoch of reionization. Independent constraints 
on the tempetarure evolution of IGM would be helpful to constrain this 
further. 

\subsection{Matter density and Hubble parameter}

\begin{figure}
\centerline{\psfig{file=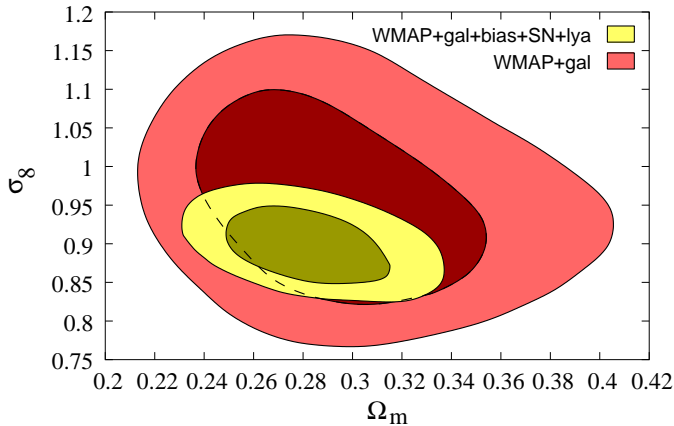,width=3.5in}}
\caption{
68\% and 95\%
contours in the $(\Omega_m, \sigma_8)$ plane
showing previous constraints from WMAP and galaxy clustering
with the new data.}
\label{fig15}
\end{figure}

The matter density parameter $\Omega_m$ has contributions from cold dark 
matter, 
baryons, and neutrinos. We assume spatially flat universe, so matter 
density $\Omega_m$ is related to dark energy density 
$\Omega_m=1-\Omega_{\lambda}$. 
As emphasized in \cite{2003Sci...299.1532B}, the 
matter density is still allowed to cover a wide range of values  
from the present data: 
in 7-parameter models with running WMAP+SDSS-gal gives
$\Omega_m=0.269^{+0.041}_{-0.033}$. WMAP+SDSS-lya gives a slightly
lower value with comparable error, $\Omega_m=0.257^{+0.055}_{-0.048}$ in 
models with running. 
Combining WMAP, SDSS-gal and SDSS-lya gives $\Omega_m=0.299^{+0.037}_{-0.032}$.
Including  the bias and SNIa and ignoring running 
brings the value to 
\be
0.282^{+0.021}_{-0.020}\;{}^{+0.043}_{-0.043}\;{}^{+0.066}_{-0.067}
\ee
which is a factor of 2 improvement over previous constraints. 
The matter 
density is correlated with $r$ and inclusion of tensors in the
parameter space slightly reduces the density parameter. 
There is a significant improvement in $(\sigma_8, \Omega_m)$ plane 
with the addition of new data (figure \ref{fig15}). 

\begin{figure}
\centerline{\psfig{file=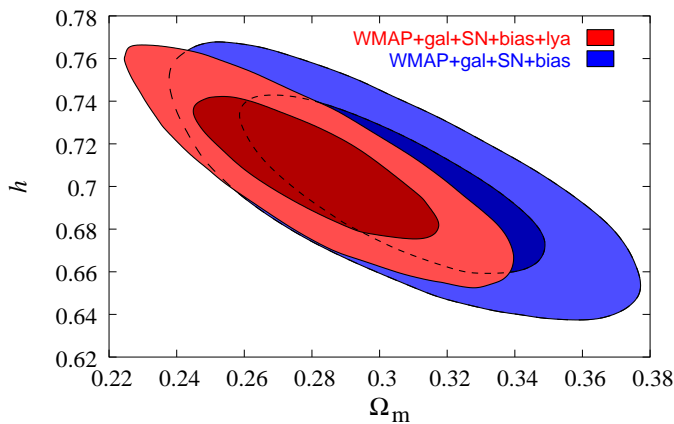,width=3.5in}}
\caption{68\% (inner, dark) and 95\% (outer, light)
contours in the $(\Omega_m,h)$ plane
with and without SDSS-lya. There is a strong 
correlation between the two parameters because WMAP constrains
best the combination $\Omega_mh^2$.
}
\label{fig13}
\end{figure}

Despite the improvements the
matter density remains strongly correlated with the Hubble parameter
$h$, as expected from the fact that $\Omega_mh^2$ is better 
determined from the CMB than each parameter separately.
This is shown in figure \ref{fig13} for 6-parameter models for 
the analysis with and without inclusion of SDSS-lya. 

For the Hubble parameter the best fit value and its error is
$h=0.71\pm 0.02$ in 6-parameter space.
In 9-parameter space with tensors, massive neutrinos 
and running we find $h=0.74\pm 0.05$.
All of these fits are 
statistically acceptable and are in 
good agreement with the HST key project 
value $h=0.72\pm 0.08$ \cite{2001ApJ...553...47F}, 
although a different group using almost the same data continues to 
find a significantly lower value $h=0.58\pm 0.06$ \cite{2003dhst.symp..222T}.


The new data also improve significantly the age of the universe 
constraint.  We find $t_0=13.6^{+0.19}_{-0.19}$Gyr, compared to $14.1^{+1.0}_{-0.9}$Gyr
found from the WMAP+SDSS-gal analysis \cite{2004PhRvD..69j3501T}.  

\subsection{Dark energy}

\begin{figure}
\centerline{\psfig{file=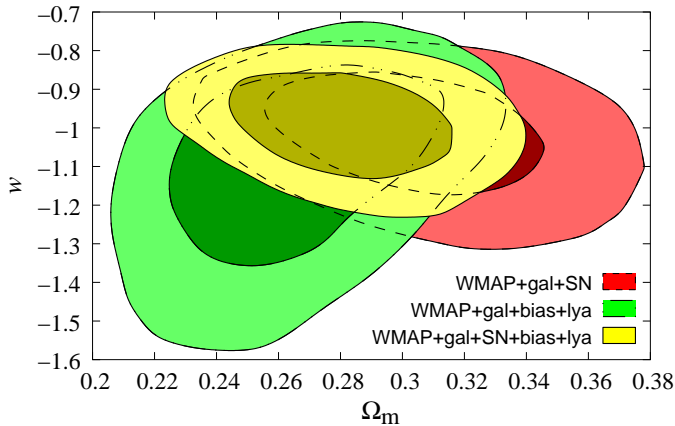,width=3.5in}}
\caption{68\% and 95\% 
contours in the $(\Omega_m, {\rm w})$ plane
showing previous constraints without SDSS-lya and bias, 
constraints without SNIa, and combined constraints. 
In all cases the data are consistent with 
a cosmological constant model (${\rm w}=-1$).
}
\label{fig3}
\end{figure}

So far we have assumed dark energy in the form of 
a cosmological constant, ${\rm w}=-1$. We now relax this 
assumption and explore the constraints on w. To maximize
the constraints we 
add to some of the analyses the ``gold" SNIa data \cite{2004ApJ...607..665R}.
Because we do not want to limit ourselves to ${\rm w}>-1$
we assume dark energy does not cluster (${\rm ndyn}=3$ option in CMBFAST4.5). 
Note that clustering of dark energy vanishes for w=-1 and so 
if w is close to -1 then it makes very little difference if 
clustering is included or not. 
Figure \ref{fig3} shows the constraints in 
the $({\rm w},\Omega_m)$ plane. We find 
\be
{\rm w}=-0.990^{+0.086}_{-0.093}\;{}^{+0.16}_{-0.201}\;{}^{+0.222}_{-0.351}.
\label{w}
\ee
We see that ${\rm w}=-1$ is an acceptable solution.  
This should be compared to ${\rm w}=-1.01^{+0.097}_{-0.12}$ 
we find 
in the absence of bias and \lyaf\ constraint, to 
${\rm w}=-0.91^{+0.13}_{-0.15}$ using the new SNIa data but  
just some of the LSS constraints 
\cite{2004astro.ph..3292W}, to 
${\rm w}=-1.02^{+0.13}_{-0.19}$ using a simple $\Omega_m$ 
prior 
\cite{2004ApJ...607..665R}, and to ${\rm w}=-0.98^{+0.12}_{-0.12}$
from the WMAP 1st year analysis \cite{2003ApJS..148..175S}. 
It is worth emphasizing the agreement and complementarity of 
the LSS, CMB, and SNIa constraints: 
in the absence of SNIa data the constraint is ${\rm w}=-1.02^{+0.15}_{-0.19}$ 
and w is positively correlated with $\Omega_m$ (figure \ref{fig3}). 
These solutions allow phantom energy models (${\rm w}<-1$) 
with w as low as -1.5 for 
low matter density values. 
On the other hand the two are anticorrelated for 
the WMAP+SDSS-gal+SNIa data constraints, and phantom energy solutions are 
allowed for high values of the matter density. 
Combing the two sets of constraints significantly reduces the 
parameter space of allowed solutions. 
All of these different combinations give very consistent results and the 
median value hardly changes at all and is in all cases very close to 
${\rm w}=-1$. 
Our constraints are a factor of 1.5-2 better than previously 
published constraints on 
the dark energy equation of state. Some of the improvement comes from 
our more sophisticated analysis which includes all of the information 
previously available and some from the new constraints from the bias and
\lyaf, which further reduce the errors. 
This is an example of how combining different data sets leads not only to a 
significant improvement in the accuracy of cosmological parameters, 
but also how consistency among the different methods 
gives confidence in the resulting constraints. 

\begin{figure}
\centerline{\psfig{file=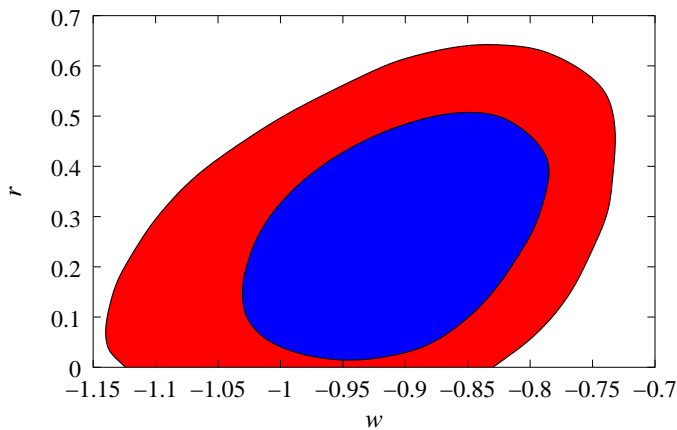,width=3.5in}}
\caption{68\% and 95\% 
contours in the $(w,r)$ plane
using the WMAP+SDSS-gal+bias+lya+SNIa  
constraints. The presence of tensors favors a slightly lower value 
of w, but the quality of the fit is only marginally improved and the  
cosmological constant model (${\rm w}=-1$) is near the 68\%
contours.
}
\label{fig11}
\end{figure}

The results are weakly model dependent, in the sense that they are
sensitive to the parameter space over which one 
is projecting. If we include tensors and running in the analysis 
we find  
\be
{\rm w}=-0.908^{+0.077}_{-0.091}\;{}^{+0.143}_{-0.197}\;{}^{+0.192}_{-0.324}~,
\ee
roughly a 1-sigma change in the central value compared to the case 
without tensors in equation \ref{w}. 
Figure \ref{fig11} 
shows that tensors and the equation of state are correlated. 
The shift in the best fitted value of w reflects a large volume of 
parameter space associated with $r>0$ models 
and not any fit improvement when adding tensors and running: $\chi^2$
changes only by 1 and there is no need to introduce tensors (or 
${\rm w} \ne -1$) to improve the fit to the data. 
We also find no correlation between the equation of state and running. 

Our constraints eliminate a significant fraction of previously 
allowed parameter space, with 95\% contours at $-1.19<w<-0.83$ without 
tensors and at $-1.11<w<-0.77$ with tensors. 
Thus a large fraction 
of the parameter space of "phantom energy" models 
with ${\rm w}<-1$ \cite{2002PhLB..545...23C} 
and tracker quintessence models with ${\rm w} \sim -0.7$ \cite{1999PhRvD..59l3504S} 
appears to be excluded. 
Other dark energy models
which predict ${\rm w}\sim -1$ remain acceptable. 
It is interesting to note that simplest quintessence solutions with ${\rm w}>-1$ are more 
acceptable if tensors are present at a level predicted by some inflationary models
($r \sim 0.2$). 

\begin{figure}
\centerline{\psfig{file=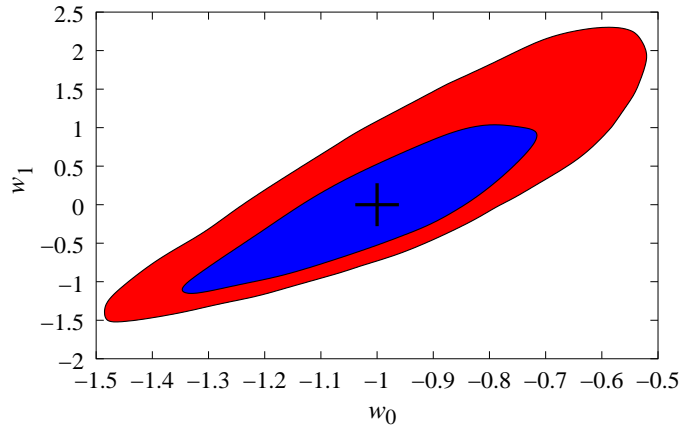,width=3.5in}}
\caption{68\% (inner, blue) and 95\% (outer, red)
contours in the $({\rm w}_0,{\rm w}_1)$ plane
using WMAP+SDSS-gal+bias+lya+SNIa measurements. 
We find that the simplest solution, ${\rm w}_0=-1$, ${\rm w}_1=0$ (marked 
by a cross),
fits the data best. 
}
\label{fig_w0_w1}
\end{figure}

We also ran a MCMC simulation exploring a non-constant equation of 
state. We use a second order expansion 
\be
{\rm w}={\rm w}_0+(a-1){\rm w}_1+(a-1)^2{\rm w}_2,
\ee
where $a=1/(1+z)$ is the expansion factor \cite{2003PhRvL..90i1301L}. 
The advantage of this expansion is that it is well behaved throughout
the history of the universe from early times, when $a\sim 0$, to today ($a=1$). 
This is in contrast to the often adopted expansion in terms of the 
redshift, ${\rm w}={\rm w}_0+{\rm w}'z$, which diverges at high redshift and so can give 
artificially tight constraints on ${\rm w}'$ if CMB (or even BBN)  
constraints at high redshift are used, without actually saying much
about the time dependence of w in the relevant regime $0<z<1$. 
In contrast, using our expansion $0<z<1$ covers half of the full range
of w
so ${\rm w}_1$ is being constrained in the 
regime of interest. If we impose ${\rm w}_2=0$ then  
the best fit values and errors we find using all the data are
\bea
{\rm w}_0&=&-0.981^{+0.193}_{-0.193}\;{}^{+0.384}_{-0.373}\;{}^{+0.568}_{-0.521} \nonumber \\
{\rm w}_1&=&0.05^{+0.83}_{-0.65}\;{}^{+1.92}_{-1.13}\;{}^{+2.88}_{-1.38}. 
\eea
We find that 
${\rm w}_0=-1$, ${\rm w}_1=0$ is well within 1-$\sigma$ contour and very close to the 
best fit model (figure \ref{fig_w0_w1}). 

\begin{figure}
\centerline{\psfig{file=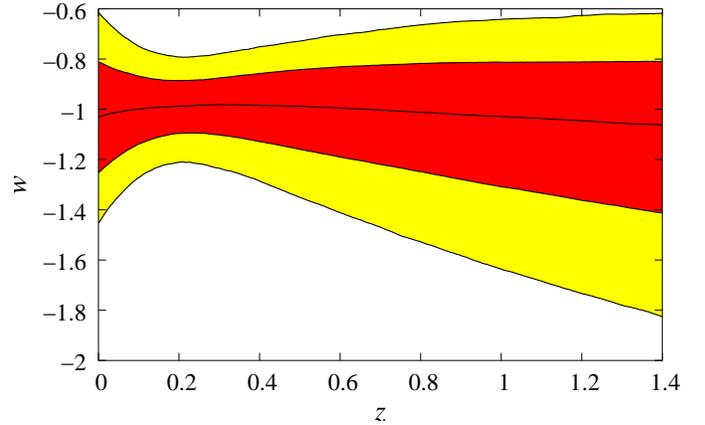,width=3.5in}}
\caption{Median (central line), 68\% (inner, red) and 95\% (outer, yellow)
intervals of ${\rm w}(z)$
using all the data in the chains without tensors and with a 
3 parameter expansion of equation of state with respect to the
expansion factor. 
Very similar results are found for the 2 parameter expansion of w, so the
constraints are reasonably model independent as long as w is a 
smooth function of redshift.
We find that the simplest solution, ${\rm w}=-1$,
fits the data at all redshifts. 
}
\label{fig_wz}
\end{figure}

The parameters ${\rm w}_0$, ${\rm w}_1$ and ${\rm w}_2$
are strongly correlated, as shown in figure \ref{fig_w0_w1} for 
the first two, so the 
error on ${\rm w}_0$ has expanded by a factor of 2 compared to the 
constant equation of state case. We can explore less model dependent 
constraints on ${\rm w}(z)$ by computing the median and 1, 2-$\sigma$
intervals from MCMC outputs at any redshift. Over a narrow range of 
redshift these contours will be nearly model independent as long as the 
equation of state is a relatively smooth function of redshift. 
We find that the data constrain best the equation of state w at $z=0.3$, 
where we find 
${\rm w}(z=0.3)=-1.011^{+0.095}_{-0.099}\;{}^{+0.176}_{-0.215}\;{}^{+0.264}_{-0.357}.$ 
Thus $z=0.3$ is the pivot point for the current measurements of equation 
of state and the constraint here is nearly model independent. 
This is confirmed by our analysis with ${\rm w}_2$. In this case 
we find severe degeneracies 
among the 3 paramaters, but the value at $z=0.3$ is 
\be
{\rm w}(z=0.3)=-0.981^{+0.106}_{-0.120}\;{}^{+0.205}_{-0.249}\;{}^{+0.269}_{-0.386}, 
\ee
which is nearly the same as for the two parameter analysis
with ${\rm w}_2=0$.  
These constraints are shown in figure \ref{fig_wz}. 

The corresponding constraint at $z=1$ for two parameter (${\rm w}_0$, 
${\rm w}_1$) analysis is 
${\rm w}(z=1)=-1.00^{+0.17}_{-0.28}\;{}^{+0.27}_{-0.66}\;{}^{+0.33}_{-1.00}$. 
Adding ${\rm w}_2$
we find 
\be
{\rm w}(z=1)=-1.03^{+0.21}_{-0.28}\;{}^{+0.39}_{-0.58}\;{}^{+0.52}_{-0.85}, 
\ee
so 1-$\sigma$ contours are nearly the same, while 2 and 3-$\sigma$ 
contours expand in the positive direction and shrink in the negative 
direction compared to 2-parameter analysis. This value is thus also relatively 
independent of parametrization. 

\begin{figure}
\centerline{\psfig{file=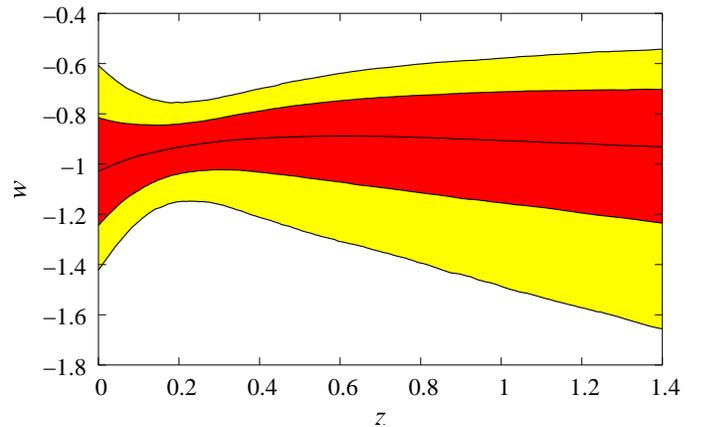,width=3.5in}}
\caption{Same as figure \ref{fig_wz} but for MCMC with tensors. 
}
\label{figwr}
\end{figure}

Adding tensors and running to the 3-parameter expanasion of w gives,
\be
{\rm w}(z=0.3)=-0.914^{+0.089}_{-0.106}\;{}^{+0.169}_{-0.225}\;{}^{+0.229}_{-0.343} 
\ee
and 
\be
{\rm w}(z=1.0)=-0.93^{+0.21}_{-0.25}\;{}^{+0.35}_{-0.56}\;{}^{+0.48}_{-0.90}. 
\ee
This is shown in figure \ref{figwr}. 
Thus, in either case, there is no evidence for any time dependence of 
the equation of state and its value is remarkably close to -1 even at 
$z=1$. As for a constant w analysis we find that 
tensors increase the preferred value of w by about 0.1. 
These constraints on the time dependence of w are significantly better 
compared to the 0.8-0.9 allowed variation between $z=0$ and 
$z=1$ found previously \cite{2004ApJ...607..665R}. 
\lyaf\ analysis measures the growth of structure in the 
range $2<z<4$ and so helps in constraining models with a significant 
component of dark energy present at $z>2$ \cite{2003MNRAS.344..776M}. 

\section{Implications for inflation}

Inflation is currently the leading paradigm for explaining the 
generation of structure in the universe. Inflation, an epoch of accelerated
expansion in the universe, explains why the universe is approximately 
homogeneous and isotropic and why it is flat \cite{1981PhRvD..23..347G,1981MNRAS.195..467S,1982PhRvL..48.1220A,1982PhLB..108..389L}. During this 
accelerated expansion quantum fluctuations are transformed into 
classical fluctuations when they cross the horizon 
(i.e., their wavelength exceeds the 
Hubble length during inflation) and can subsequently 
be observed as perturbations in 
the gravitational metric \cite{1981JETPL..33..532M,1982PhRvL..49.1110G,1983PhRvD..28..679B,1982PhLB..115..295H,1982PhLB..117..175S}. 
A generic prediction of a single field inflation models is that
the perturbations are adiabatic (meaning that all the species in
the universe are unperturbed
on large scales except for the overall shift caused by the 
perturbation in the metric) and Gaussian. These
predictions, together with flatness ($K=0$),
have been explicitly assumed in our analysis. 

We note here that cyclic/ekpyrotic models \cite{2002Sci...296.1436S} are 
an alternative to inflation, which, 
despite a very different starting point and without a period of
accelerated expansion, lead to almost identical 
predictions as inflation \cite{2003PhRvL..91p1301K}. Specifically, these models 
predict no observable tensor contribution, 
spectral index 
$n_s$ close to unity, and negligible running \cite{2001PhRvD..64l3522K}. 
Very specific forms of
cyclic potentials have not been explored in much detail
in these models and for this reason we will not discuss them 
explicitly below, but most of our constraints on the form of the inflationary 
potential can easily be translated 
into the corresponding constraints on the form of cyclic model potential. 

Here we will explore a class of single field inflation models, 
in which there is a single field responsible for the dynamics of 
inflation (even though additional fields may be present or even required
to end inflation, as in the case of hybrid inflation \cite{1994PhRvD..49..748L}). 
We will assume the early universe is dominated by a minimally
coupled scalar field $\phi$, which we will express in Planck mass
units setting $8\pi G=1$.
During inflation 
the energy density is dominated by potential $V$.
The Hubble
parameter $H^2=V/3$ is nearly constant and the equation of state
is ${\rm w}=p/\rho\sim -1$. Since $H=d\ln a/dt$ it follows that 
the expansion factor is exponentially increasing with time, 
$a=a_{\rm end}e^{H(t-t_{\rm end})}$. One can introduce the 
number of e-folds before the end of inflation at time $t_0$ as 
\be
N=\ln (a_{\rm end}/a_0)=
\int_{t_0}^{t{\rm end}} H(t)dt=\int_{\phi_0}^{\phi_{\rm end}}{V \over V'}d\phi, 
\ee
which can be computed for any specific form of the potential. 
Here we will define it to be the number of e-folds before the end of 
inflation when the pivot point, $k_{\rm pivot}=0.05/$Mpc, crosses the horizon. 
Note that the usual definition is with respect to the 
largest observable scale, $k \sim 10^{-3}$/Mpc, which corresponds
to  $\Delta N =4$ larger number of efolds.
The latter number is expected to be between 50-60 efolds for standard inflation 
(64 for $V\propto \phi^4$), 
but could be as low as 20 or as high as 100 in special cases 
\cite{2003PhRvD..68j3503L,2003PhRvL..91m1301D}. 
For our pivot point choice
we will thus adopt $N=50$ as the standard value (60 for $V\propto \phi^4$), 
but also explore more general constraints on it. 

If the kinetic energy density were negligible all the time the universe 
would keep
exponentially expanding and there would be no end to inflation. Typically
therefore one must have deviations from the pure ${\rm w}=-1$ case. These 
deviations lead not only to a finite number of efolds, but also break the 
scale invariance of the primordial power spectrum.  
Since we know from current observational constraints 
that $r<1$ and $n_s\sim 1$ we can 
adopt the slow-roll approximation 
to relate the form of the potential to the 
observed quantities $r$, $n_s$, $\alpha_s$, and $\Delta^2_{\cal R}$.
The slow-roll parameters are defined as \cite{2000cils.conf.....L}
\begin{eqnarray}
\epsilon_V &=&{1\over 2}\left({V'\over V}\right)^2
\nonumber \\
\eta_V &=&  {V''\over V}
\nonumber \\
\xi_V &=&{V'V'''\over V^2}.
\end{eqnarray}
Note that in some early literature the 3rd slow-roll parameter $\xi$ was
denoted as $\xi^2$ to emphasize the point that it is generically 
of second order in $\epsilon$ or $\eta$ \cite{1995PhRvD..52.1739K}. 
We will not use this 
notation since $\xi$ can be positive or negative and since it does not 
have to be of second order in the slow-roll expansion. 

The relations between the slow-roll parameters and observables are
\begin{eqnarray}
\Delta^2_{\cal R} & = & { V \over 24\pi^2 \epsilon_V } \nonumber \\
r & = & 16\epsilon_V \nonumber \\
n_s -1 & = & -6\epsilon_V+2\eta_V \nonumber \\
\alpha_s & =& 16\epsilon_V \eta_V -24\epsilon_V^2-2\xi_V.
\label{slowroll}
\end{eqnarray}
As mentioned in the previous section, we assume $r=-8n_T$ and do not 
consider the running of the tensor spectral index, both of 
which should be valid for single field inflation in the relevant regime. 

\begin{figure}
\centerline{\psfig{file=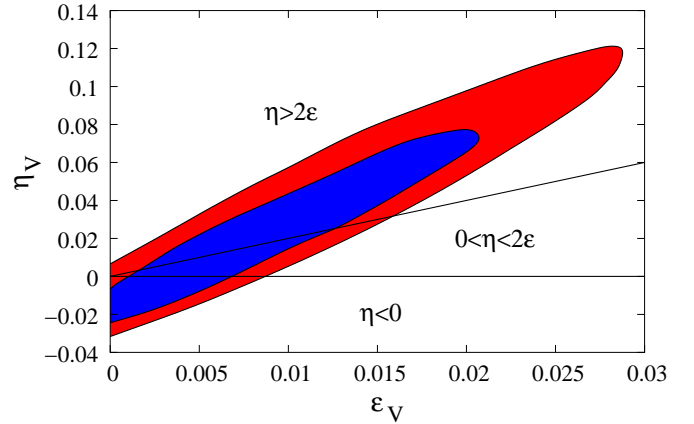,width=3.5in}}
\caption{68\% (inner, blue) and 95\% (outer, red) 
contours in the $(\epsilon_V,\eta_V)$ plane
using all the measurements without running (table 1, 5th column). 
Also shown are the regions occupied by the 
3 classes of inflationary models. 
All 3 classes of models are allowed, but 
individual models within each class are constrained. 
Note that the solutions disfavor low energy models ($\epsilon_V=0$)
with large positive curvature ($\eta_V>0$), typical of hybrid inflation
models, as well as models where both $\epsilon_V$ is large 
and $\eta_V<\epsilon_V/2$, 
typical of chaotic inflation models with steep potentials.  
}
\label{fig12}
\end{figure}

\begin{figure}
\centerline{\psfig{file=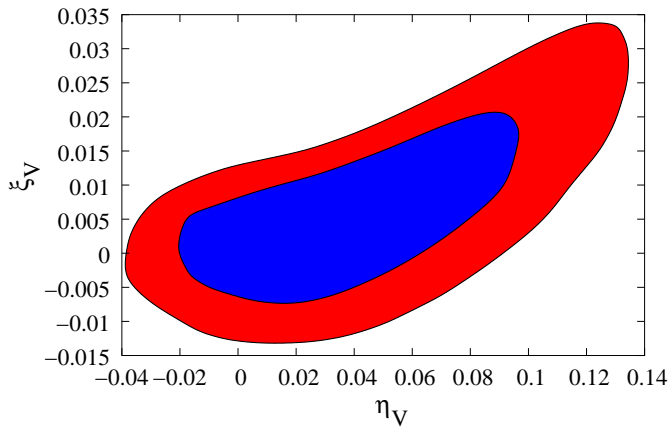,width=3.5in}}
\caption{68\% (inner, blue) and 95\% (outer, red) 
contours in the $(\eta_V,\xi_V)$ plane
from MCMC with running and tensors (table 2, 5th column). 
}
\label{fig14}
\end{figure}

Traditionally the inflationary models are divided into separate classes
depending on the value of first two slow-roll parameters \cite{1997PhRvD..56.3207D,1999PhR...314....1L,2000cils.conf.....L}. 
Figure \ref{fig12} shows the distribution in the $(\epsilon_V,\eta_V)$ plane. 
We see that both  positive and negative values of $\eta$ are allowed and that 
there is a strong correlation between the two from the observational 
constraints, a consequence of positive correlation between tensors and 
primordial slope.
Figure \ref{fig14} shows the distribution in the $(\eta_V,\xi_V)$ plane. 
Both parameters are consistent with 0. 
The basic constraints are 
$\epsilon < 0.03$, $-0.04<\eta <0.12$ and $-0.015-<\xi_V<0.035$, 
so all slow roll
parameters are small.

\subsection{Large field models}

The simplest inflationary models 
are the monomial potentials, $V =V_0 \phi^p$, for which 
the first two parameters are comparable, $\epsilon \sim \eta$, and the
curvature is positive, $\eta>0$. These potentials occur in chaotic 
inflation models \cite{1983JETPL..38..176L}. 
In these models a deviation from scale invariance, $n_s-1 =-(2+p)/2N$,
also implies a significant tensor contribution, $r=4p/N$, while 
running is negligible, $\alpha_s=-2(n_s-1)^2/(p+2)=-(p+2)/2N^2$. 
Because both slow-roll parameters are of order $(p/\phi)^2$
these chaotic inflation-type 
potentials require a large field, $\phi>1$, 
to satisfy observationally required 
$r<1$ and $n_s \sim 1$. For this reason these 
models are sometimes called large field models. While this may limit their 
particle physics motivation there are brane inspired 
models where this property can be justified \cite{2003PhRvL..90v1302A}. 
More generic parametrization of these models in terms 
of curvature is $0<\eta_V< 2\epsilon_V$. 

\begin{figure}
\centerline{\psfig{file=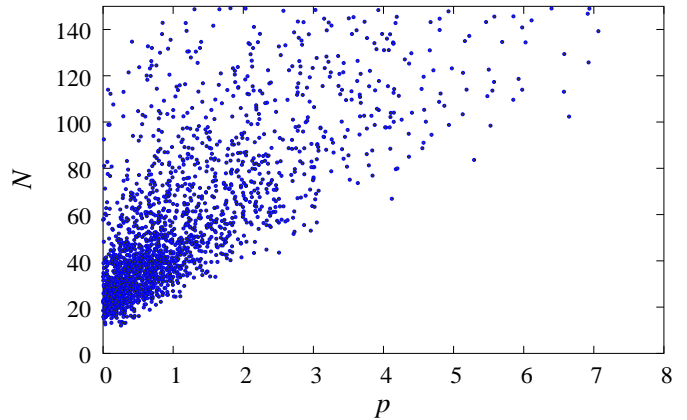,width=3.5in}}
\caption{Scatter plot of MCMC solutions with $n_s<1$ converted into 
the $(p,N)$ plane assuming relations valid for chaotic inflation models. 
Here $p$ is the slope of the inflationary potential and $N$ is the 
number of e-folds.  For $N<60$ we require $p<3$.  }
\label{fig10}
\end{figure}

With the exception of $p=2$, 
chaotic models are not particularly favored from our analysis. Figure
\ref{fig6} shows the position in the $(r,n_s)$ plane for two representative 
cases, $p=2$ and $p=4$.  
We find that the
$V \propto \phi^2$ model ($n_s=0.96$, $r=0.16$ for N=50) 
is within the 2-sigma contour, while 
the $V \propto \phi^4$ model ($n_s=0.95$, $r=0.27$ for N=60) 
is outside the 3-sigma contour, since 
it predicts more tensors and a redder spectrum for that tensor amplitude 
than observed. Figure 
\ref{fig10} shows all chain elements with $n_s<1$ converted to $(p,N)$
values using the expressions above. For standard inflation we require $N<60$
and this limits us to $p<3$. Similarly, 
figure \ref{fig12} shows that $\epsilon_V > \eta_V/2$ with large $\epsilon_V$
models are disfavored. 

For specific models we also minimized $\chi^2$ by exploring 
all of the parameter space of the remaining parameters and compared 
that to the global minimum in $\chi^2$. 
We find $\Delta \chi^2=5$ for the $V \propto \phi^2$ model and 
$\Delta \chi^2=13$ for the $V \propto \phi^4$ model. 
These results are in agreement 
with the MCMC results and show that the latter case is excluded at more than 
$3-\sigma$ confidence.

\subsection{Large positive curvature models}

We turn next to models with positive large curvature, $\eta>2\epsilon$. 
A generic potential of this type can be obtained by adding a 
constant to the monomial potential, $V=V_0(1+c\phi^p)$, where $c$ 
is a positive dimensionless constant. 
These models allow small field solutions to inflation, $\phi\ll 1$, and 
so are popular for model building in the context of supersymmetry. 
In this limit, and if dimensionless $c$ is not too large, one has 
$\epsilon \ll 1$. 
In such models, inflation never ends (since the potential never drops to zero), 
so another field must be brought in to accomplish this. 
Hybrid inflation is an example of such a mechanism \cite{1994PhRvD..49..748L}.
If $\epsilon$ is small then
these models predict $r \sim 0$ and $n_s>1$ (equations \ref{slowroll}, 
the latter condition requires $\epsilon < \eta/3$). 
For $p=2$ the slope is constant, $n_s-1=2c$ and
there is no running, 
while for $p>2$ running is negative
and is given by $\alpha_s=-(p-2)/(p-1)[(n_s-1)]^2/2$. This is always small 
since a large deviation in the $n_s>1$ direction is strongly disfavored, 
so $\alpha_s\sim 0$. 
Some of these models are disfavored: for $r=0$ and in the absence of 
running we find $n_s<1.0$ at 90\% confidence and $n_s<1.04$ at 99.9\% 
confidence, so 
if $\epsilon_V \sim 0$ then 
$\eta_V>0.02$ is excluded at 3 sigma. 
Thus the deviations from scale invariance have to be very small
for these models to be acceptable. 

\subsection{Large negative curvature models}
The most promising models from the observational perspective are negative 
curvature models, $\eta < 0$. 
As noted above, the main reason that large positive curvature models are
disfavored is that in the absence of tensors the 
data favor $n_s<1$, while small positive curvature models are disfavored
because they predict large tensors and a red spectrum at the same time, 
whereas the data are more consistent with blue spectrum if tensors are 
significant. 
A generic potential of negative curvature models
can be obtained by switching the sign
on the hybrid potential form, $V=V_0(1-c\phi^p)$, where $c$ 
is a positive dimensionless constant. In these models the field $\phi$ is 
slowly rolling from low to high values until reaching the point 
where the potential vanishes at $c\phi^p=1$, at which point inflation stops.
This is a generic scenario of spontaneous symmetry breaking models as in
the first working inflation model, that of new inflation \cite{1982PhRvL..48.1220A}. 
For $p=2$ the slope is again constant at $n_s-1=-2c$ and there is no running. 

In these models one has
$n_s-1=-2(p-1)/(p-2)/N$ and $\alpha_s=-(p-2)/(p-1)[(n_s-1)]^2/2$. 
The running is of order $(n_s-1)^2/2$ and the prefactor is 
unity at best, so running is negligible. The slope $n_s$ ranges 
between 0.96 (in the limit of $|p|\rightarrow \infty$,where $n_s-1=-2/N$) 
and 1, in excellent agreement with observational constraints. 

One finds good agreement using other potentials proposed in the 
literature, such as the potential based on 
one-loop correction in a spontaneous symmetry broken SUSY \cite{1994PhRvL..73.1886D}. 
The potential is of the form $V=V_0[1+\alpha \ln (\phi/Q)]$. In this 
model the number of e-folds is of the order $N=\phi^2/2\alpha$ (this 
expression works best if $\alpha \ll 1$). This model predicts 
$n_s-1=-2\alpha [1+3\alpha/2]/\phi^2$ and $\alpha_s=-(n_s-1)^2[2\alpha+3\alpha^2/2+1/2]/[1+3\alpha/2]^2$. Running is again negligible. 
Solutions with $\phi\ll 1$ require $\alpha\ll \phi^2\ll 1$,
in which case
the slope becomes $n_s-1=-1/N=-0.02$ for $N=50$, in 
excellent agreement with the observed value $n_s=0.971^{+0.023}_{-0.019}$. 

Many other models in this class also work. 
A model often mentioned as an example of allowing a large running is the 
softly broken SUSY model with $V=V_0(1-c\phi^2(\ln(\phi/\phi_*)-1/2)/2$. 
This model has a large 3rd derivative for small field $\phi$, 
$V'''/V_0=-c/\phi$, 
so it can lead to large $\xi$ and large runnings. 
For this model 
there is an inequality relation between 
slope and running of the form
$\alpha_s>-{(n_s-1)^2 \over 4}>-2\times 10^{-3}$,
so a large negative running cannot be accommodated in this model for 
the allowed values of $n_s$. 
Our solutions do not favor large negative runnings anyways, unless
one is willing to consider models with massive neutrinos whose 
mass exceeds 0.3eV, so this model is acceptable, but it can overpredict 
the running on the positive side. 

There are also examples of
models which can change from one inflationary case to the other, 
such as hybrid model with one-loop correction \cite{1997PhRvD..56.1841L}, 
$V=V_0\left[1+\alpha \left(\ln (\phi/Q) +{c \over 4}\left({\phi \over \phi_0}\right)^p\right)\right]$, 
which under specially arranged
 conditions causes the slope to change 
from $n_s>1$ on large scale to $n_s<1$ on small scale. 
Again, there is no evidence for such a transition 
in the data, so there is no need to consider these special cases. 

Finally, there are models that predict the simplest possible 
case of $r=0$, $n_s=1$ and $\alpha_s=0$ \cite{2004JCAP...04..001A}. These models 
are perfectly acceptable from our data. 

While we only surveyed a small subset of inflationary models here, 
it is clear that their generic prediction is a nearly 
scale invariant spectrum, $|n_s-1|<0.05$, little or no 
tensors, $r<1$ and small running, $\alpha_s \sim 10^{-3}$. 
All of these predictions agree with our constraints. 
Running is 
a particularly powerful test of standard inflationary (and cyclic) models 
in the sense that if running turned out to be large, a large class of 
inflationary models would have been eliminated. 
The original suggestions of running in the WMAP data sparked a lot 
of theoretical interest in inflationary models with  
running \cite{2003PhRvD..68b3508K,2003PhRvD..68f3501C}, but such
models are unnatural in the sense that they 
require a feature in the potential at exactly the scale of observations 
today. Our results suggest that the natural prediction of inflation, 
small running, is confirmed by observations. 

\section{Conclusions}
In this paper we performed a joint cosmological analysis of 
WMAP, the SDSS galaxy power spectrum and its bias, the SDSS \lyaf\
power spectrum, and the latest supernovae SNIa sample. 
We work in the context of current structure formation models, such as
inflation or cyclic models, so we assume spatially flat universe and 
adiabatic initial conditions.
We also ignore more exotic components such as warm dark matter.
The new ingredients, 
SDSS \lyaf\ and SDSS bias, lead to a significant reduction 
of the errors on all the parameters. 
Many parameters are improved in 
accuracy by factors of two or more.  
For example, for the amplitude of fluctuations
we find $\sigma_8=0.90\pm 0.03$ and for the matter
density we find $\Omega_m=0.28\pm 0.02$, both a significant 
improvement over previous constraints. From the fundamental physics 
perspective the highlights of the new constraints are: 

1) The scale invariant primordial power 
spectrum is a remarkably good fit to the data and there is no 
evidence that the spectral index deviates from the scale 
invariant value $n_s=1$, nor is there any evidence of its running 
with scale. We also find no evidence of tensors in the joint analysis. 
The constraints on 
running have improved 
by a factor of 3 compared to an analysis without  
the new \lyaf\ constraints.
These provide a
data point at $2<z<4$ and $k \sim 1$/Mpc, a significantly smaller scale
than scales traced by the CMB and galaxies. 

2) There is no cosmological evidence of neutrino mass yet.
In the standard models with 3 neutrino families we find for
the total neutrino mass $\sum m_{\nu}<0.42$eV (95\% c.l.). 
When our 
analysis is combined with atmospheric and solar neutrino 
experiments \cite{2004hep.ex..4034A,2001PhRvL..87g1301A} we 
find that neutrino masses are not degenerate: the 
most massive neutrino family has to be at least 10\% more massive than 
the least massive family, $m_3/m_1>1.1$: 
the mass of the least massive neutrino family has to be 
$m_{1}<0.13$eV, and that of the most massive neutrino family 
$m_3<0.15$eV, both at 95 \% c.l.  
In alternative models with a 4th massive neutrino family in addition 
to 3 (nearly) massless ones we find $m_{\nu}<0.79$eV, excluding 
all of the allowed LSND islands at 95\% c.l.

3)  Dark energy continues to be best characterized as a standard
cosmological constant with constant energy density and equation 
of state ${\rm w}=-1$. 
When all the data is combined together the error on w is 0.09,
a reduction compared to previously published 
values \cite{2003ApJS..148..175S,2004astro.ph..3292W,2004ApJ...607..665R}. 
A cosmological constant with ${\rm w}=-1$ is remarkably 
close to the best fit value for a variety of different subsamples of 
the data. A significant region of phantom energy parameter space with 
${\rm w}<-1$ is excluded, as are some of the tracker quintessence 
models with ${\rm w} \sim -0.7$. 
The current data do not support any time dependence of the
equation of state.

As the statistical 
errors are being reduced the required level at which systematics 
must be controlled increases as well. Our limits on cosmological 
parameters assume that the errors from the SDSS \lyaf\, 
SDSS power spectrum shape, SDSS bias, 
WMAP CMB power spectrum, 
and the SNIa data are all properly characterized by the authors
and that there are no additional sources of systematic error. 
Each one of these ingredients has to 
be tested and redundancy
is necessary for the results to be 
believable. In our extensive tests we find no evidence of a 
disagreement between the different observational inputs, but 
further tests with these and other data sets 
are needed to verify and confirm our results. 
In addition, the upcoming 2 year analysis of WMAP polarization will 
improve the constraints on the optical depth and reduce the 
errors on parameters correlated with it. 

Tests of the basic model are
particularly important for \lyaf\ , which is responsible 
for most of the improvement on the primordial power spectrum shape 
and amplitude. Despite the extensive tests presented in \cite{McDonald04c}, 
more 
work is needed to investigate all possible physical effects
that can modify its distribution and to see how these may affect 
the conclusions reached in this paper.
Some of these tests will come from the ongoing work on SDSS data, 
such as the bispectrum analysis. 
Similarly, more work is needed to verify the accuracy of 
simulations with independent hydrodynamic codes. 
The present analysis, together with its sister papers
\cite{McDonald04b,McDonald04c},
is not the final word on this subject, but merely a first attempt
to take advantage of the enormous increase in statistical power 
given by the SDSS data \cite{2004astro.ph..5013M}. 
Current analysis marginalizes over many physical processes that have 
little or no external constraints and as a result the statistical 
power of cosmological constraints from the \lyaf\ 
is weakened. 
Better theoretical understanding 
of these processes together with external constraints from additional 
observational tests could lead to a significant reduction 
of observational errors on the primordial slope and its running 
even with no additional improvements in the observations. 

In summary, adding SDSS \lyaf\ and SDSS bias constraints to 
cosmological parameter estimation leads to 
a significant improvement in the precision with which the cosmological 
parameters can be determined. 
Despite these improvements
we find no surprises. 
Many of these results are not unexpected, but the tightness of 
the constraints is rapidly eliminating many of the alternative 
models of structure formation, neutrinos and dark energy. 
Future cosmological observations and improvements in theoretical 
modelling will allow us
to verify the constraints found here and improve them further. 
As the constraints become tighter there may be additional surprises
awaiting us in the future. 

We thank the WMAP and SNIa teams for creating the data sets used 
in present analysis. 
Our MCMC simulations were run on a Beowulf cluster at Princeton University, 
supported in part by NSF grant AST-0216105.
US is supported by a fellowship from the 
David and Lucile Packard Foundation, 
NASA grants NAG5-1993, NASA NAG5-11489 and NSF grant CAREER-0132953.

Funding for the creation and distribution of the SDSS Archive has been provided by the Alfred P. Sloan Foundation, the Participating Institutions, the National Aeronautics and Space Administration, the National Science Foundation, the U.S. Department of Energy, the Japanese Monbukagakusho, and the Max Planck Society. The SDSS Web site is http://www.sdss.org/.

The SDSS is managed by the Astrophysical Research Consortium (ARC) for the Participating Institutions. The Participating Institutions are The University of Chicago, Fermilab, the Institute for Advanced Study, the Japan Participation Group, The Johns Hopkins University, Los Alamos National Laboratory, the Max-Planck-Institute for Astronomy (MPIA), the Max-Planck-Institute for Astrophysics (MPA), New Mexico State University, University of Pittsburgh, Princeton University, the United States Naval Observatory, and the University of Washington.

\bibliography{apjmnemonic,cosmo,cosmo_preprints}

\end{document}